\documentclass[12pt,preprint]{aastex}

\usepackage{amsmath,amssymb,latexsym,graphics}

\shorttitle{Microlensing in WFI\,2033--4723}
\shortauthors{Morgan et al.}

\newcommand{\kms}{\textrm{km\,s}^{-1}}
\newcommand{\mpc}{\textrm{Mpc}}
\newcommand{\msun}{M_{\sun}}

\begin{document}

\title{Accretion Disk Size Measurement and Time Delays in the Lensed Quasar WFI\,2033--4723}

\author{Christopher W.\ Morgan\altaffilmark{1}, 
Gregory E.\ Hyer\altaffilmark{1}, Vivien Bonvin\altaffilmark{2}, Ana M. Mosquera\altaffilmark{1}, Matthew Cornachione\altaffilmark{1}, 
Frederic Courbin\altaffilmark{2}, Christopher S. Kochanek\altaffilmark{3,4} and Emilio E.\ Falco\altaffilmark{5} }

\altaffiltext{1}{Department of Physics, United States Naval Academy,
572C Holloway Rd, Annapolis, MD 21402, USA; cmorgan@usna.edu}
\altaffiltext{2}{Laboratoire d'Astrophysique, \'{E}cole Polytechnique F\'{e}d\'{e}rale de Lausanne (EPFL),
Observatoire, 1290 Sauverny, Switzerland}
\altaffiltext{3}{Department of Astronomy, The Ohio State University, Columbus, OH 43210, USA}
\altaffiltext{4}{Center for Cosmology and Astroparticle Physics, The Ohio State University, Columbus,
OH 43210, USA}
\altaffiltext{5}{Harvard-Smithsonian Center for Astrophysics, 60 Garden St, Cambridge, MA 02138, USA}

\begin{abstract}

We present 13 seasons of $R$-band photometry of the quadruply-lensed quasar WFI\,2033-4723
from the 1.3m SMARTS telescope at CTIO and the 1.2m Euler Swiss Telescope at La Silla, in which we detect microlensing variability of $\sim0.2$~mags on 
a timescale of $\sim$6~years.  Using a Bayesian Monte Carlo technique, we analyze the microlensing signal to obtain a measurement of the 
size of this system's accretion disk of $\log (r_s/{\rm cm}) = 15.86^{+0.25}_{-0.27}$ at $\lambda_{rest} = 2481{\rm \AA}$, assuming a $60^\circ$ 
inclination angle.   We confirm previous 
measurements of the BC and AB time delays, and we obtain a tentative measurement of the delay between the closely spaced A1 and A2 images of 
$\Delta t_{A1A2} = t_{A1} - t_{A2} = -3.9^{+3.4}_{-2.2}$~days. We conclude with an update to the Quasar Accretion Disk Size -- Black Hole Mass Relation, in which we confirm that
the accretion disk size predictions from simple thin disk theory are too small.
 \end{abstract}

\keywords{gravitational lensing: strong --- gravitational lensing: micro ---
                    accretion disks --- quasars: individual (WFI\,2033-4723)}

\section{INTRODUCTION}\label{sec:intro}

Gravitationally lensed quasars provide a wealth of resources to observers.  Their utility in cosmology was realized quite early on \citep[e.g.][]{Refsdal1964}, and 
a number of collaborations (e.g. {\it COSMOGRAIL}, \citet{courbin05} \& {\it H0LiCOW}\footnote{http://www.h0licow.org}) continue to pursue measurements of lensed quasar time delays 
to make independent measurements of the Hubble Constant, $H_0$ \citep[e.g.][]{kochanek02,Vuissoz2008,fohlmeister2013,Suyu2017,bonvin17} 
and a range of other useful cosmographic measurements. Quasar microlensing, 
also predicted quite some time ago \citep[e.g.][]{Chang1979}, provides additional motivation for monitoring lensed quasars since the analysis of microlensing variability \citep[e.g.][]{Kochanek2004, morgan06, poindexter07, morgan10, hainline13, macleod15} and chromatic flux ratio anomalies \citep[e.g.][]{pooley07,bate08,bate11,mediavilla11,pooley12,schechter14}
can be analyzed to probe the central engines of the quasars and the properties of the lens galaxy.  

\begin{figure}
\epsscale{1.0}
\plottwo{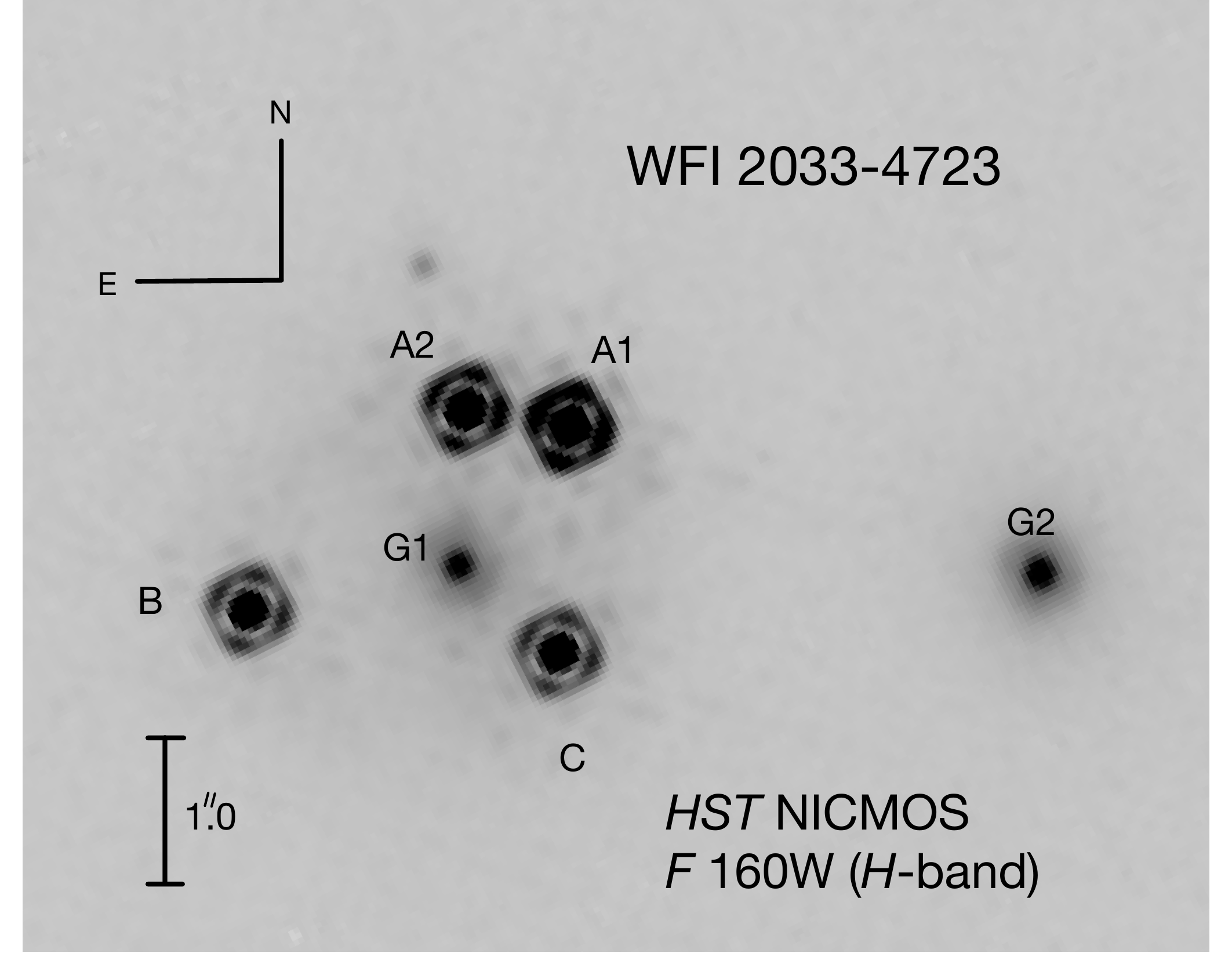}{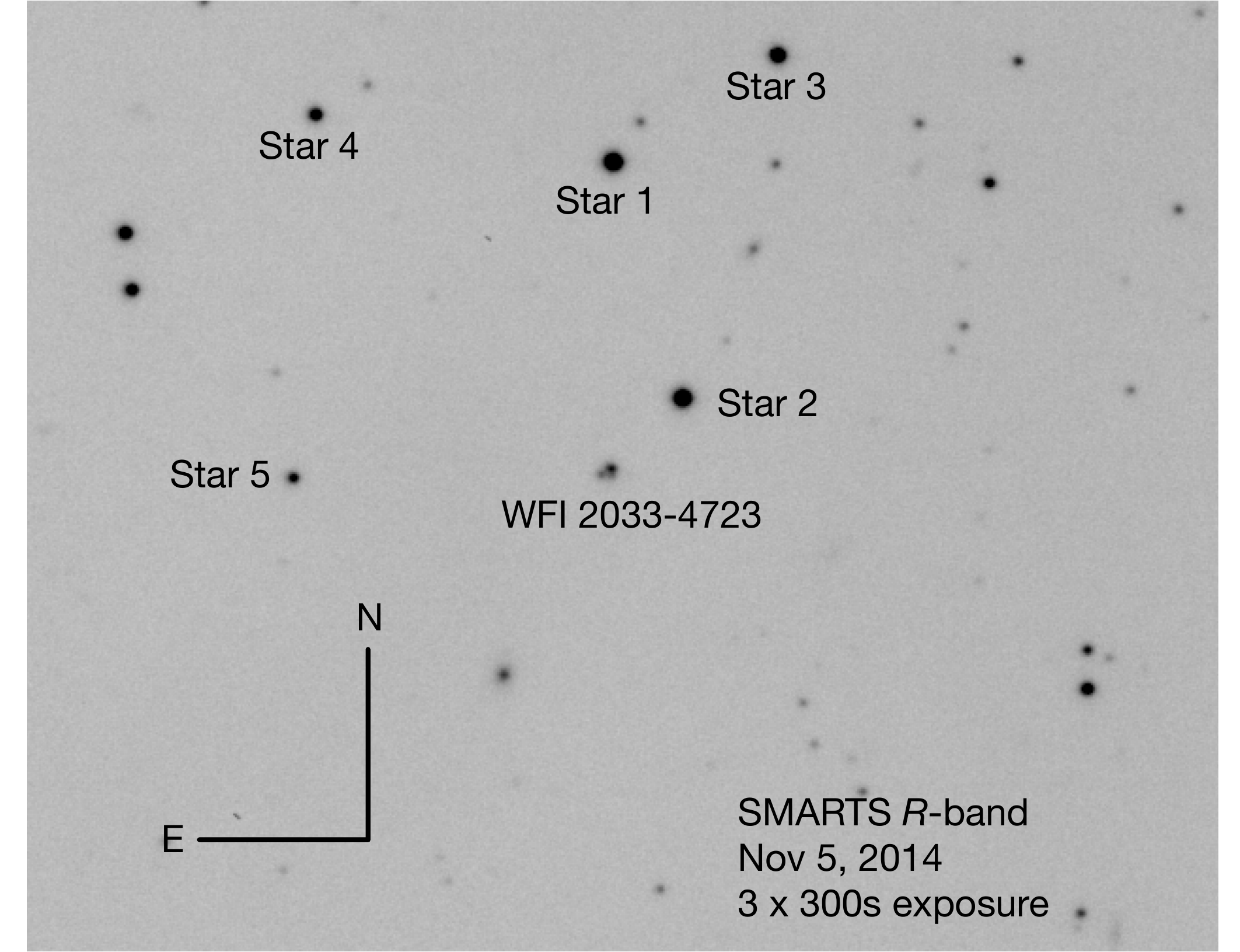}
\caption{Left: $HST$ $H$-band NICMOS image of the lensed quasar WFI 2033--4723.  Images A1 and A2 are merging at a fold caustic; we expect them to have very similar time delays.
Right: A stack of three 300~s $R$-band images from the 1.3m SMARTS telescope. The image scale is $4' 33" \times 3'39"$.
\label{fig:hst}}
\label{fig:HST}
\end{figure}

The quadruply lensed quasar WFI J2033-4723 (hereafter WFI2033; 20$^{h}$33$^{m}$42.08$^{s}$, -47$^{\circ}$23'43.0" [J2000.0]) 
was discovered during a wide-field imaging survey for lensed quasars in the southern hemisphere using the MPG/ESO 2.2 m 
telescope \citep{morgan04}. It has a source redshift of $z_s$ = 1.66, a 
lens redshift of $z_l=0.661$ \citep{Eigenbrod06} and a maximum image separation of $2\farcs5$. 
\citet[][hereafter V08]{Vuissoz2008} used three seasons (2004-2007) of monitoring data from the Small and Moderate Aperture Research Telescope System (SMARTS) 
1.3m telescope at the Cerro Tololo Inter-American Observatory (CTIO) and the 1.2m Leonhard EULER Swiss telescope 
at La Silla, Chile to measure time delays of $|\Delta t_{B-A}| = 35.5 \pm 1.4$ days and $|\Delta t_{B-C}| = 62.6^{+4.1}_{-2.3}$ days between 
the merged A1+A2 = A, B and C images. 
V08 found no evidence of variability due to extrinsic factors such 
as microlensing.  Recently, however, \citet{Giannini2017} made a robust detection of microlensing in their 4-season monitoring campaign using the 
1.54m Danish telescope at La Silla, a result which we independently corroborated in this investigation.  Most recently, \citet{Motta2017} used the single-epoch chromatic 
microlensing technique to make estimates of the size of the central engine and broad line region in WFI2033.

In this paper, we combine 9 new seasons of WFI2033 monitoring data with the 4 seasons of data from V08 to create a 13-season set of light curves.  We 
 present our observational data and reduced light curves in \S\ref{sec:obs_data}, and we analyze the full combined light curves in \S\ref{sec:delays} to 
 confirm the V08 time delays and measure the A1--A2 delay for the first time. In \S\ref{sec:analysis} we describe our microlensing analysis technique to include the properties of our
 strong lens models for WFI2033.  In \S\ref{sec:results}, we present the results of our analysis, and we discuss their implications for accretion disk theory.
Throughout our discussion we assume a flat cosmology with $\Omega_{\textrm{M}}=0.3$, $\Omega_{\Lambda}=0.7$, and $H_{0}=70\,\kms~\mpc^{-1}$ \citep{hinshaw09}.

\section{OBSERVATIONAL DATA}\label{sec:obs_data}

\subsection{$HST$ Imagery}\label{sec:HST}
We observed WFI2033 in the $V$- (F555W), $I$- (F814W) and $H$- (F160W) bands using the {\it Hubble Space Telescope} ({\it HST}) as an element of the the CfA-Arizona 
Space Telescope Survey  \citep[CASTLES\footnote{http://cfa.harvard.edu/castles/},][]{lehar00}.  The $V$- and $I$-band images were taken using the 
Wide-Field Planetary Camera 2 (WFPC2). 
The $H-$band images, originally presented in V08, were taken using the Near-Infrared Camera and Multi-Object
Spectrograph (NICMOS). We fit the astrometry and photometry of the lens in the {\it HST} imagery 
with the {\it imfitfits} \citep{lehar00} routine, using a de Vaucouleurs model for the lens galaxy G1, 
an exponential disk model for the quasar host galaxy and point sources for the quasar images.   
Our astrometric and photometric fits, consistent with those made independently by V08, are presented in Table~\ref{tab:HST}.

\begin{deluxetable}{cccccc}
\tablecaption{HST Astrometry and Photometry of WFI 2033--4723}
\tablehead{Component &\multicolumn{2}{c}{Astrometry}
                 &\multicolumn{3}{c}{Photometry}\\
                 \colhead{}
                 &\colhead{$\Delta\hbox{RA}$}
                 &\colhead{$\Delta\hbox{Dec}$}
                 &\colhead{H=F160W}
                 &\colhead{I=F814W}
                 &\colhead{V=F555W}
                 }

\startdata
A1  &$-2\farcs196\pm0\farcs003$ &$1\farcs261\pm0\farcs003$ &$17.22\pm0.02$ &$18.15\pm0.05$ &$19.24\pm0.03$\\
A2  &$-1\farcs482\pm0\farcs003$ &$1\farcs376\pm0\farcs003$ &$17.60\pm0.02$ &$18.65\pm0.13$ &$19.26\pm0.09$\\
B  &$\equiv 0$ &$\equiv 0$ &$17.85\pm0.02$ &$18.63\pm0.14$ &$19.24\pm0.04$\\
C  &$-2\farcs114\pm0\farcs003$ &$-0\farcs277\pm0\farcs003$ &$17.90\pm0.02$ &$18.82\pm0.03$ &$19.53\pm0.02$\\
G1  &$-1\farcs438\pm0\farcs006$ &$0\farcs308\pm0\farcs009$ &$17.46\pm0.00$ &$19.69\pm0.03$ &$22.51\pm0.28$\\
\enddata
\label{tab:HST}
\end{deluxetable}

\subsection{Monitoring Observations}\label{sec:monitoring}
On the 1.3 m SMARTS telescope we used the optical channel of the dual-beam ANDICAM instrument \citep{depoy03},which has a plate scale of 0\farcs{369}~pixel$^{-1}$ and a $6\farcm5 \times 6\farcm3$ field of view. The mean sampling of the SMARTS data is one epoch every eight days, with three 300~s exposures at each epoch using the \textit{R}-band filter. The \textit{R}-band filter has an effective wavelength of 658 nm, translating to a rest-frame wavelength of  2473~\AA \, in the UV. 
The SMARTS dataset consists of 117 epochs between March 2004 and August 2017.

\begin{figure}
\plotone{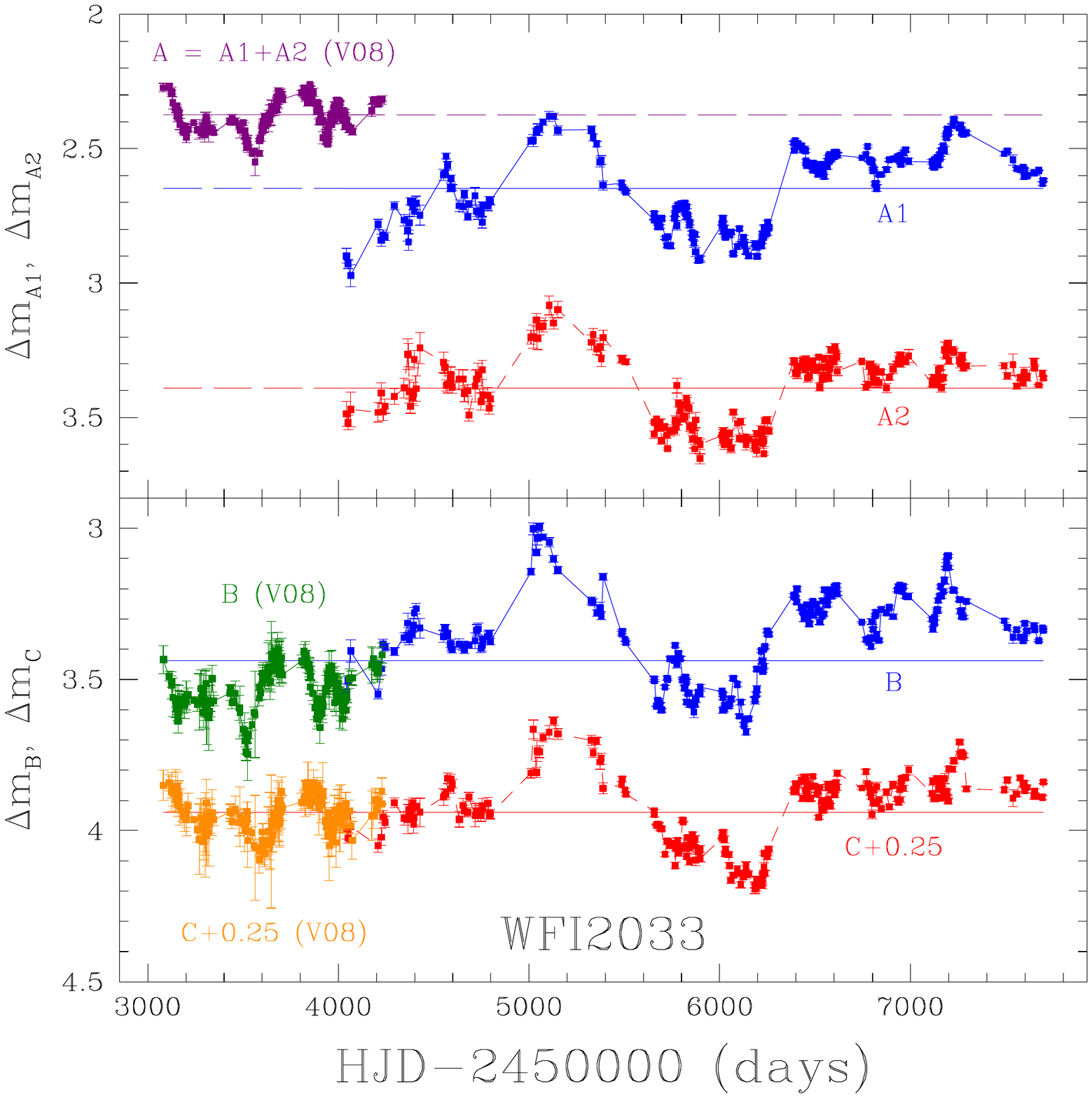}
\vskip -72pt
\caption{Combined light curves from WFI2033. Images A1 and A2 are shown in the top panel and images B and C are plotted in the bottom panel.
The previously published light curves from \citet{Vuissoz2008} are labeled with `V08', and in those light curves the flux from images A1 and A2 were summed. 
Magnitudes are relative to an arbitrary zero point.}
\label{fig:lc}
\end{figure}

On the 1.2 m EULER telescope we used the EulerCAM camera which has a plate scale of 0\farcs{2149}~pixel$^{-1}$ and a $15\farcm0 \times 15\farcm0$ field of view. The mean sampling of the EULER data is one epoch every five days, with five 360s exposures at each epoch using the $`RG'$ or `Rouge Gen\`{e}ve' filter. The $RG$ filter is a modified broad Gunn \textit{R} filter, with an effective wavelength of 660 nm translating to a rest-frame wavelength of 2481~\AA. The new EULER dataset consists of 178 epochs between October 2010 and December 2016.

The details of our photometric measurement technique are 
discussed in \citet{kochanek06}, but we provide a brief summary of that process here. We use five reference stars, located at $(-2\farcs4, 61\farcs4)$, 
$(-16\farcs1, 15\farcs3)$, $(56\farcs0, 70\farcs3)$, $(60\farcs2, -0\farcs9)$ and $(-34\farcs4, 82\farcs5)$, with respect to image A1.  These reference stars are 
used as the basis for a  
three-component elliptical Gaussian point-spread function (PSF) model, which we apply to the blended quasar images to obtain the relative brightness of each component 
at each epoch.  
When applying the PSF model, we hold the relative positions of the lens components fixed to the astrometry from our \emph{HST} $H$-band images.  
We model the lens galaxy using a nested Gaussian with fixed effective radius and flux to approximate a de Vaucouleurs profile. For the effective radius, we used
our measurement from the $HST$ fits $r_{eff} = 0\farcs83 \pm 0\farcs1$, and for the flux we use the
value which minimizes the total $\chi^{2}$ in the residuals following galaxy model subtraction when summed over all epochs. 
We measured a very small color offset of 0.002 magnitudes between the EULER and SMARTS photometry which we applied to the
SMARTS data when creating our combined light curves and the data provided in Tables~\ref{tab:lcsmarts} and \ref{tab:lceuler}.
Both the EULER and SMARTS images
are characterized by a median stellar FWHM (seeing) of $1\farcs2$.  Since the merging A1/A2 pair are separated by only $0\farcs72$, deconvolving the  
flux from these two images was challenging. For seeing conditions worse than $1\farcs5$ and $1\farcs62$ for SMARTS and EULER, respectively, we were unable to reliably resolve any of the 
quasar's images, so we were forced to discard images taken under these conditions.  We also discarded 31 of the 326 total observing 
epochs from SMARTS and EULER due to bright sky or
cloudy observing conditions.  
In Figure~\ref{fig:lc} we display our new light curves alongside the published light curves from V08. Since V08 were unable to reliably separate the 
flux from images A1 and A2, they summed the flux from this closely spaced merging pair to create a single image A light curve in which $f_A = f_{A1} + f_{A2}$.

\section{TIME DELAYS}\label{sec:delays}
\begin{figure}
\plotone{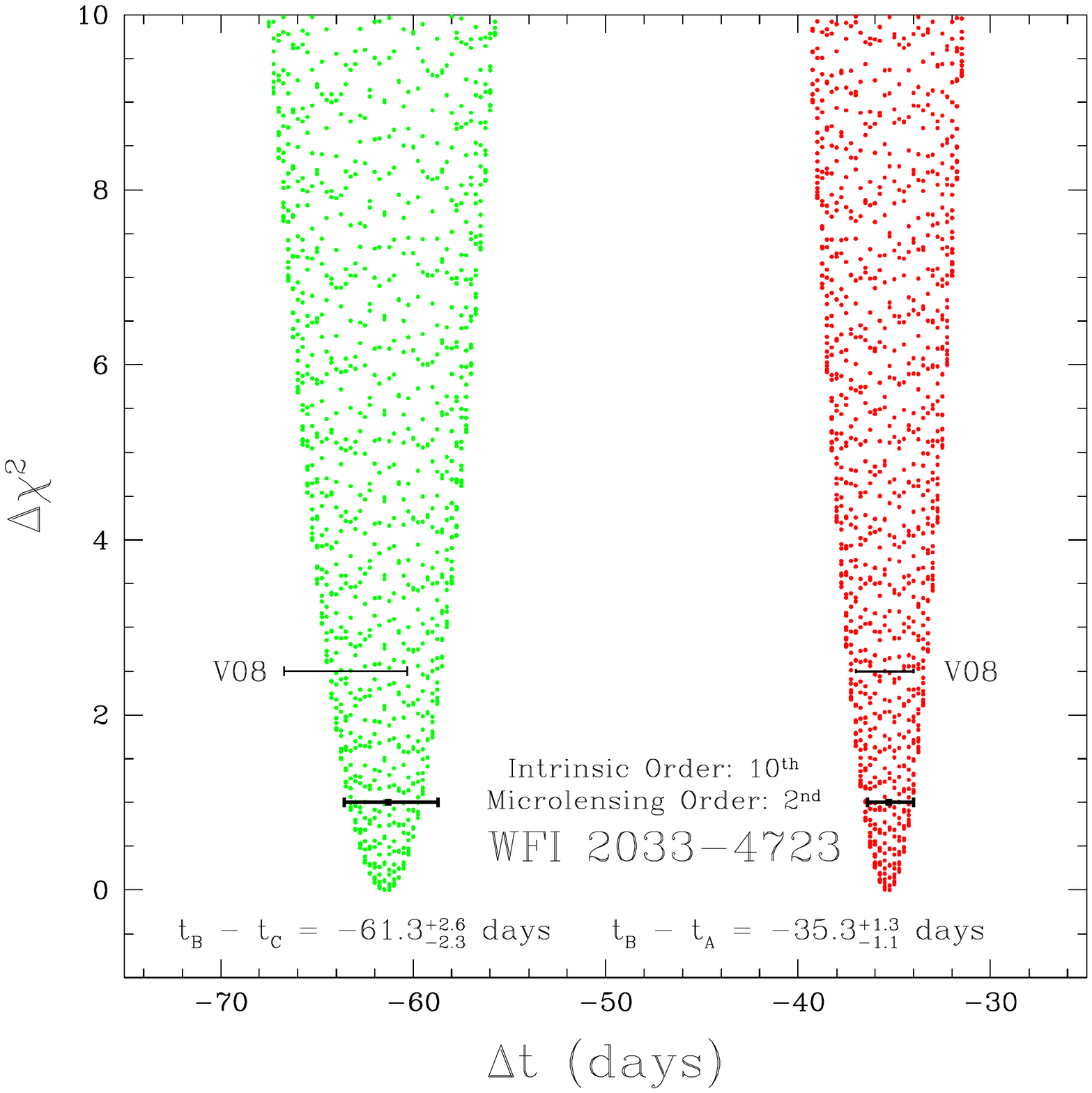}
\vskip -72pt
\caption{Left: $\chi^2$ surfaces for our measurement of the time delay between images B and A = A1 + A2 and between images B and C. 
Delays from V08 are labeled as such, and the present results with $1\sigma$ uncertainties are plotted as heavy solid points with error bars. \label{fig:bcdelay}}
\end{figure}

\begin{figure}
\vskip -72pt
\plotone{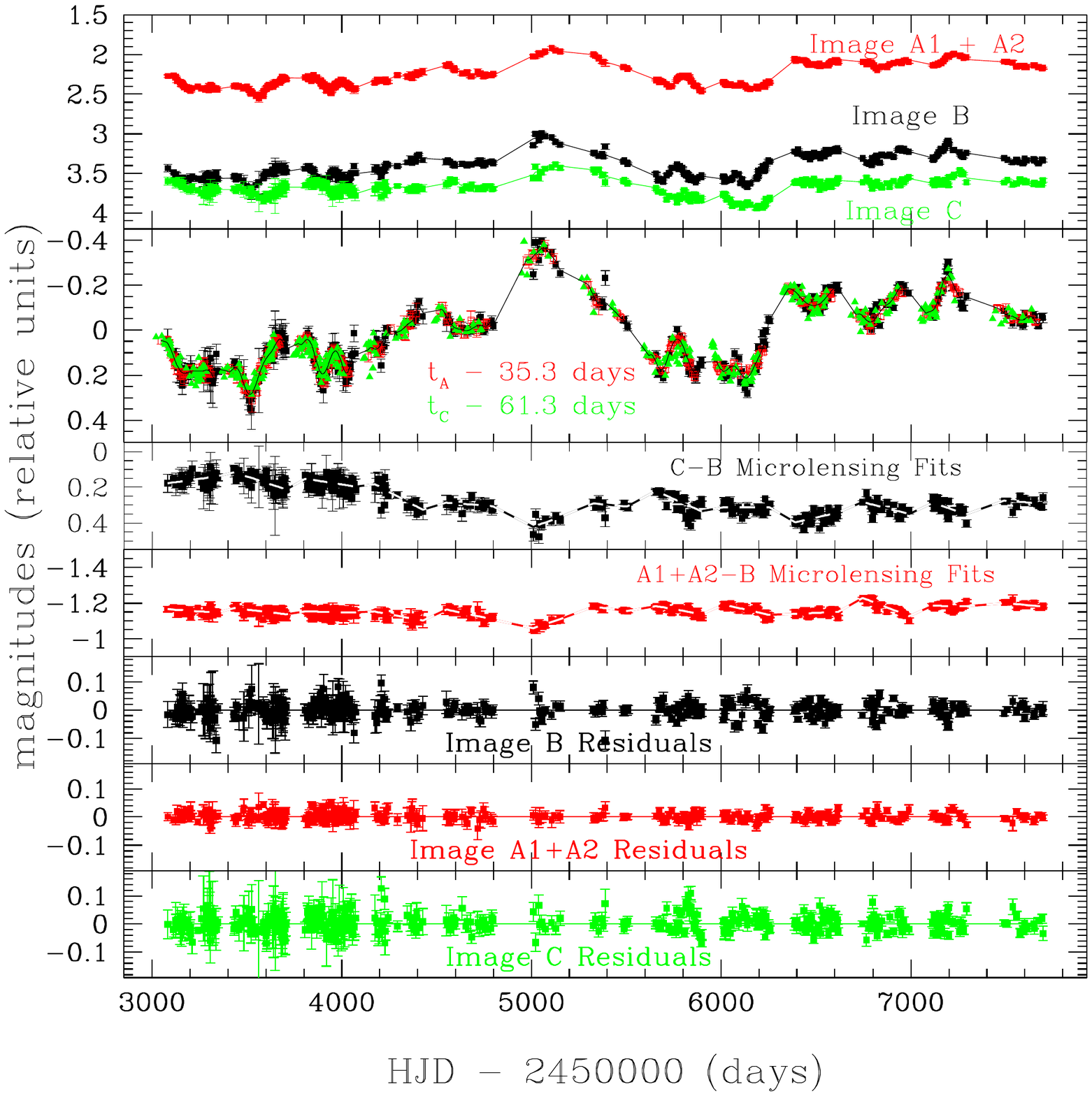}
\vskip -90pt
\caption{Top panel shows the raw light curves for A=A1+A2, B \& C. Second panel shows time-delay shifted 
light curves in which the $2^{\rm{nd}}$ order phased polynomial fit to the microlensing variability has been removed. Third and fourth panels show
the $2^{\rm{nd}}$ order fit to the microlensing in the difference light curves C-B and A-B. Note that the microlensing residuals in the B-C fits are significantly 
larger than those in the A-C fits.  The light curves were also normalized
to the same magnitude scale for display purposes.
The solid line shows the  $10^{\rm{th}}$ order polynomial fit of the 
intrinsic variability. The bottom panels show the residuals following subtraction of both the intrinsic and microlensing fits. \label{fig:bcdelayfits}}
\end{figure}

Using the polynomial light curve fitting technique of \citet{kochanek06}, we measured the time delays between the combined image A = A1 + A2, image B and image C. 
In the \citet{kochanek06} method, the intrinsic and extrinsic variability in the light curves are fitted by Legendre polynomials, and the polynomial order is chosen using the F-test.
In the case of the delay between images A and B and images C and B, we found that a $N_{source}=10^{{\rm th}}$ order polynomial provided a sufficient fit for the source variability
and that a $N_{\mu}=2^{{\rm nd}}$ order polynomial was appropriate for approximating and removing the microlensing variability. In Figure~\ref{fig:bcdelay}, we show the $\chi^2$ statistic 
for the time delay fits. The delay measurements and their $1\sigma$ uncertainties are 
$\Delta t_{BA} = t_B - t_A = -35.3^{+1.3}_{-1.1}$~days and $\Delta t_{BC} = t_B - t_C = -61.3^{+2.6}_{-2.3}$~days (in the 
sense that image B leads both images A and C). In Figure~\ref{fig:bcdelayfits}, we show the light curves shifted by these delay values.   
These new measurements are fully consistent with those of V08.

\begin{figure}
\plottwo{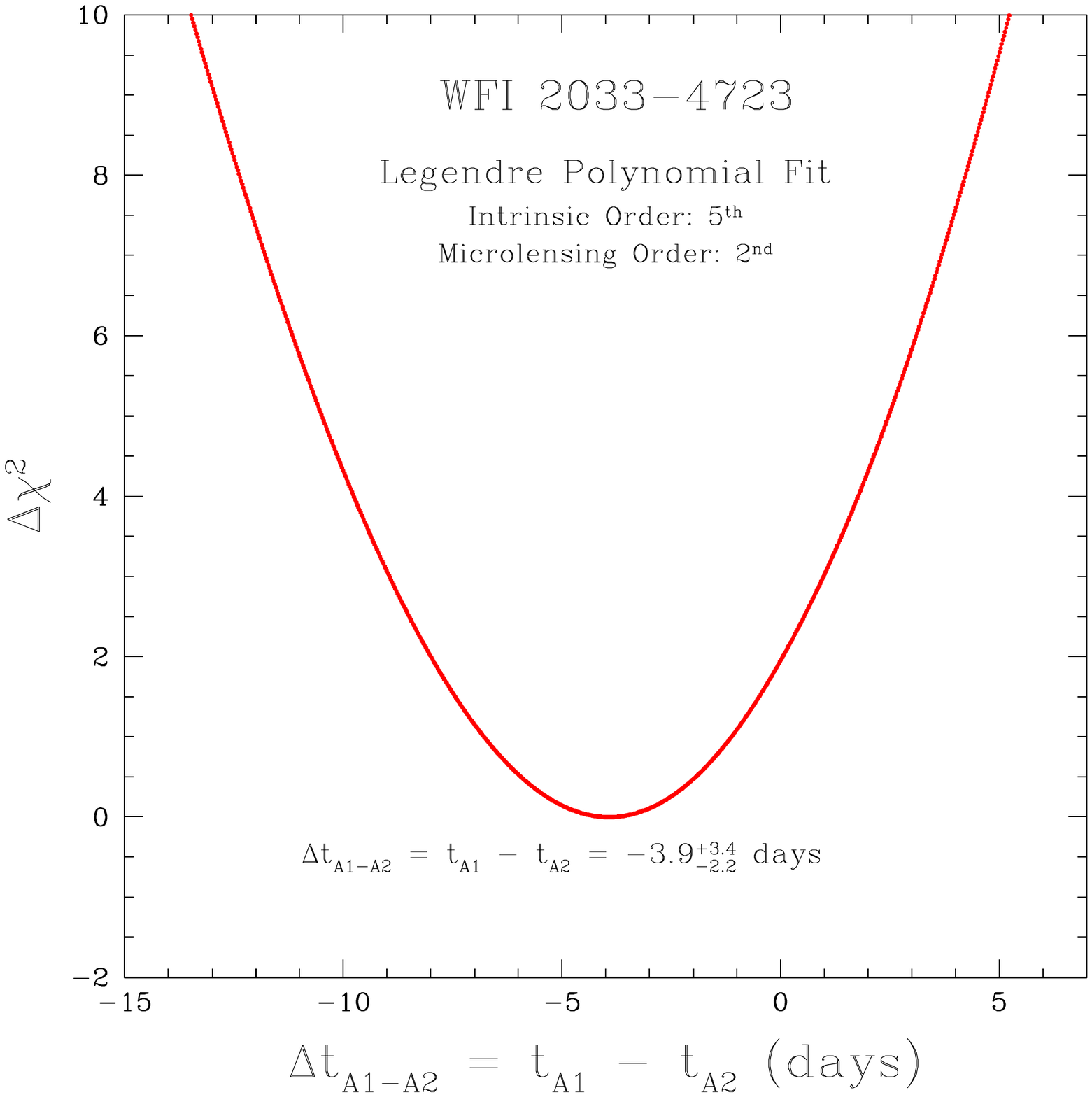}{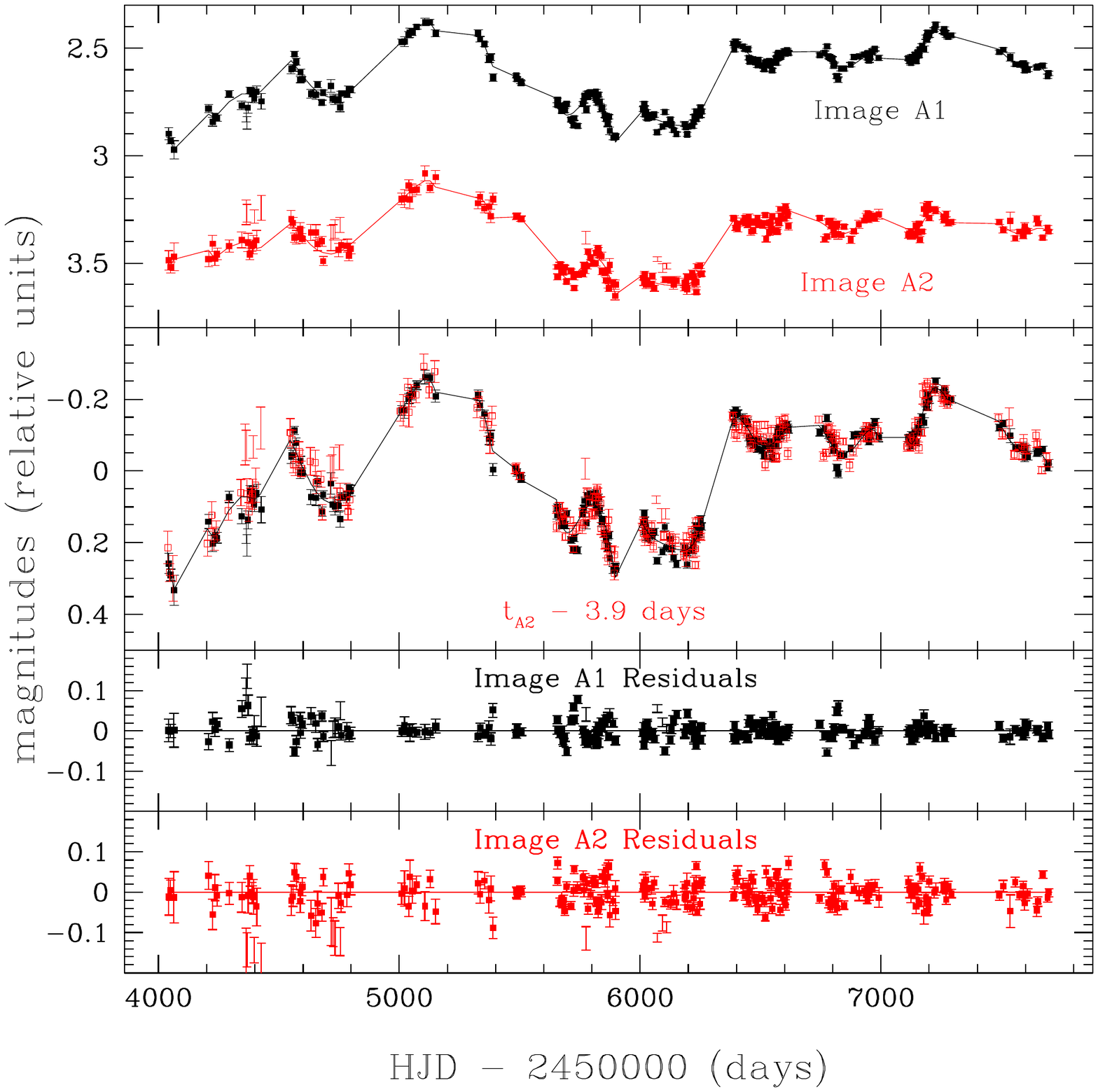}
\caption{Left: $\chi^2$ surfaces for our measurement of the time delay between images A1 and A2.  Right: Top panel shows the raw light curves. Middle panel shows time-delay shifted and
normalized 
light curves in which the $2^{\rm{nd}}$ order polynomial fit to the microlensing variability has been removed.  The solid line shows the  $5^{\rm{th}}$ order best fit to the 
intrinsic variability. Bottom panel shows residuals following subtraction of both the intrinsic and microlensing fits. \label{fig:a1a2delay}}
\end{figure}

Using our newly reduced light curves, we also obtain a tentative measurement of the delay between images A1 and A2.  With a $N_{source}=5^{{\rm th}}$ and 
$N_{\mu}=2^{{\rm nd}}$ order polynomials for the intrinsic and microlensing variability, respectively, we find that image A1 leads image A2 by 
$\Delta t_{A1A2} = t_{A1} - t_{A2} = -3.9^{+3.4}_{-2.2}$~days.   We display these results in Figure~\ref{fig:a1a2delay}. While V08 were not able to measure the A1-A2 delay, they 
did constrain the expected range of that delay to $-1 > \Delta t_{A1A2:model} > -3$~days using a series of lens galaxy mass models.  While significantly coarser 
than our measurement of the B-A and B-C delays, the A1-A2 measurement is consistent with the V08 lens models, although this pair will have the largest fractional uncertainties from
microlensing-induced variability \citep{tie18}.  In the present paper, we generate a series of lens galaxy models in which 
the expected A1-A2 delay is $-1.6 > \Delta t_{A1A2(model)} > -3.3$~days, also consistent with our A1--A2 measurement.  With these updated time delay measurements, we 
proceed to the primary goal of this investigation, the analysis of extrinsic variability from microlensing in the reduced light curves.  A full analysis of the updated delays will be
published in \citet{Bonvin2018}.

\section{MICROLENSING ANALYSIS}\label{sec:analysis}

\subsection{Lens Galaxy Models and Magnification Patterns}
In essence, our Bayesian Monte Carlo technique for microlensing analysis is an attempt to reproduce the observed microlensing variability using a large set of 
models for the physical conditions that might have led to this variability \citep{Kochanek2004}.  All of this hinges on our ability to 
accurately model the conditions in the lens galaxy through which the quasar's light must pass. 
We started by applying the LENSMODEL code of \cite{Keeton2001} to the astrometry from our $HST$ observations to yield a range of 
 models for the stellar and dark matter content in the lens galaxy at the positions of the lensed images.
Following V08, we adopted a 2-component model for the lens galaxy (G1 in Fig. \ref{fig:HST}).  Since this system is now known to exhibit microlensing of 
both the continuum and the Broad Line Region (BLR) \citep{sluse12,Motta2017}, we did not use the $HST$ flux ratios 
or those from our ground-based observations as a constraint on the lens galaxy mass models.  We required 
our models to reproduce the astrometry of the lensed images, and we allowed the position, effective radius, ellipticity and position angle 
of the lens galaxy to vary within the uncertainties of the 
photometric model presented in \S\ref{sec:HST}.  Consistent with V08, we were unable to model the astrometry of the lensed images 
unless we included the influence of the neighboring galaxy G2, the east-west shear from which cannot be created by G1 since it has an ellipticity position angle of 
only $25\degr$  east of north.  We modeled G2 as a singular isothermal sphere whose properties were also allowed to vary within the uncertainties of our $HST$ photometric
and astrometric fits.

Since the dark matter content in the lens galaxy is unknown, we created a series of 10 models for the lens galaxy
in which the dark matter fraction varies across an order of magnitude.  We
began by modeling the lens galaxy using only a de Vaucouleurs profile.  In each subsequent model, we decreased the monopole moment of the 
stellar de Vaucouleurs component by 10\% of the constant $M/L$ model mass, and we added an extended, concentric \citet[][NFW]{navarro96} 
component to model the dark matter.  We parameterized this series using the quantity
$f_{M/L}$, representing the fraction of the lens galaxy mass relative to the constant mass-to-light (M/L) ratio model.   
From this model sequence, we extract the total convergence $\kappa$,
the convergence from the stars $\kappa_*$, the shear $\gamma$ and the shear position angle $\theta_{\gamma}$ at the location of each lensed image.  
While models in the range $0.4 \le f_{M/L} \le 0.5$ are more consistent with our measured time delays, for completeness we use the 
entire model sequence in our Monte Carlo microlensing
simulations because \citet{schechter2002} demonstrated that local microlensing statistics are very sensitive to $\kappa_*/\kappa$.  The parameters of all 10 models
are presented in Table~\ref{tab:models}.   

\begin{deluxetable}{cccccccccccccc}
\tablewidth{0pt}
\tablecaption{WFI2033--4723 Lens Galaxy Mass Models}
\tablehead{\colhead{$f_{M/L}$}
                &\multicolumn{4}{c}{Convergence $\kappa$}
                &\multicolumn{4}{c}{Shear $\gamma$}
                &\multicolumn{4}{c}{$\kappa_{*}/\kappa$}
                &\colhead{$\chi^2/N_{dof}$}\\
                \colhead{}
                &\colhead{A1}
                &\colhead{A2}
                &\colhead{B}
                &\colhead{C}
                &\colhead{A1}
                &\colhead{A2}
                &\colhead{B}
                &\colhead{C}
                &\colhead{A1}
                &\colhead{A2}
                &\colhead{B}
                &\colhead{C}
                &\colhead{}
                          }
\startdata
$0.1$ &$ 0.75 $&$ 0.83 $&$ 0.61 $&$ 0.92 $&$ 0.13 $&$ 0.30 $&$ 0.15 $&$ 0.27 $&$ 0.014 $&$ 0.025 $&$ 0.010 $&$ 0.031 $&$ 0.247$ \\
$0.2$ &$ 0.71 $&$ 0.80 $&$ 0.57 $&$ 0.89 $&$ 0.15 $&$ 0.35 $&$ 0.15 $&$ 0.33 $&$ 0.029 $&$ 0.051 $&$ 0.021 $&$ 0.063 $&$ 0.431$ \\
$0.3$ &$ 0.68 $&$ 0.76 $&$ 0.53 $&$ 0.85 $&$ 0.17 $&$ 0.40 $&$ 0.15 $&$ 0.39 $&$ 0.046 $&$ 0.079 $&$ 0.033 $&$ 0.097 $&$ 0.606$ \\
$0.4$ &$ 0.65 $&$ 0.72 $&$ 0.50 $&$ 0.82 $&$ 0.18 $&$ 0.45 $&$ 0.16 $&$ 0.45 $&$ 0.07 $&$ 0.11 $&$ 0.05 $&$ 0.14 $&$ 0.761$ \\
$0.5$ &$ 0.61 $&$ 0.68 $&$ 0.45 $&$ 0.78 $&$ 0.20 $&$ 0.51 $&$ 0.16 $&$ 0.51 $&$ 0.09 $&$ 0.16 $&$ 0.07 $&$ 0.19 $&$ 0.904$ \\
$0.6$ &$ 0.58 $&$ 0.64 $&$ 0.42 $&$ 0.75 $&$ 0.22 $&$ 0.56 $&$ 0.17 $&$ 0.57 $&$ 0.11 $&$ 0.19 $&$ 0.08 $&$ 0.22 $&$ 1.02$ \\
$0.7$ &$ 0.54 $&$ 0.60 $&$ 0.37 $&$ 0.70 $&$ 0.24 $&$ 0.63 $&$ 0.17 $&$ 0.64 $&$ 0.16 $&$ 0.27 $&$ 0.13 $&$ 0.30 $&$ 1.15$ \\
$0.8$ &$ 0.50 $&$ 0.56 $&$ 0.33 $&$ 0.67 $&$ 0.26 $&$ 0.68 $&$ 0.18 $&$ 0.69 $&$ 0.19 $&$ 0.32 $&$ 0.16 $&$ 0.35 $&$ 1.24$ \\
$0.9$ &$ 0.45 $&$ 0.51 $&$ 0.27 $&$ 0.63 $&$ 0.28 $&$ 0.76 $&$ 0.19 $&$ 0.77 $&$ 0.29 $&$ 0.47 $&$ 0.27 $&$ 0.49 $&$ 1.39$ \\
$1.0$ &$ 0.43 $&$ 0.48 $&$ 0.26 $&$ 0.60 $&$ 0.29 $&$ 0.78 $&$ 0.19 $&$ 0.81 $&$ 0.27 $&$ 0.46 $&$ 0.25 $&$ 0.48 $&$ 1.41$ \\

\enddata
\tablecomments{Convergence $\kappa$, shear $\gamma$ and the fraction of the total surface density 
composed of stars $\kappa_{*}/\kappa$ at each image location for the series of 
mass models. The parameter $f_{M/L}=1.0$ is the mass of the de Vaucouleurs model for the visible lens galaxy relative to its mass in the absence of dark matter.
The $\chi^2$ per degree of freedom for each model is provided in the $\chi^2/N_{dof}$ column. }
\label{tab:models}
\end{deluxetable}

Using the inverse ray-shooting technique \citep[as described in][]{kochanek06}, we generated forty random realizations of the expected microlensing magnification conditions in the
vicinity of each image for each of our 10 macro models parameterized by $f_{M/L}$. The magnification patterns are $8192 \times 8192$ 
pixels representing a projected source plane scale of 
twenty $1 M_{\sun}$ Einstein radii or $8.66 \times 10^{17}$~cm.  This implies a pixel scale of $1.06\times10^{14}$~cm in the source plane.  We assumed an initial stellar mass function (IMF) of 
$dN(M)/dM \propto M^{-1.3}$ with a dynamic range 
of 50, which approximates the microlensing-based Galactic bulge IMF of \citet{gould00}, although \citet{wyithe00} and \citet{congdon07} show that 
microlensing statistics are not especially sensitive to choice of IMF.

\subsection{Monte Carlo Method} \label{sec:monte}

Armed with 400 sets of magnification patterns for a range of lens galaxy mass models, we used the Monte Carlo light curve fitting technique of \citet{Kochanek2004} 
as modified by \citet{morgan08b} to model unresolved image pairs.  
Since the light curves are a full 13 seasons in length, we binned them using a window of $\delta t_{bin} = 20$~days to reduce computation time.
The twenty day binning window was sufficiently short to avoid overly smoothing the microlensing variability while adequately reducing
the run time for the Monte Carlo routine. The date of each twenty-day bin was set as the mean Heliocentric 
Julian Date (HJD) of the measurements included in that bin.  
Since the light curves from V08 do not provide individual measurements for the A1 and A2 images, we adjusted the statistical weight of the A = A1+A2 data points 
to appropriately account for the combined fluxes in these cases.  Also, for the combined cases we used either the A1 or A2 magnification pattern (with equal frequency) when 
creating a simulated light curve for the combined A1 + A2 image.  
 
\begin{figure}
\epsscale{1.0}
\plottwo{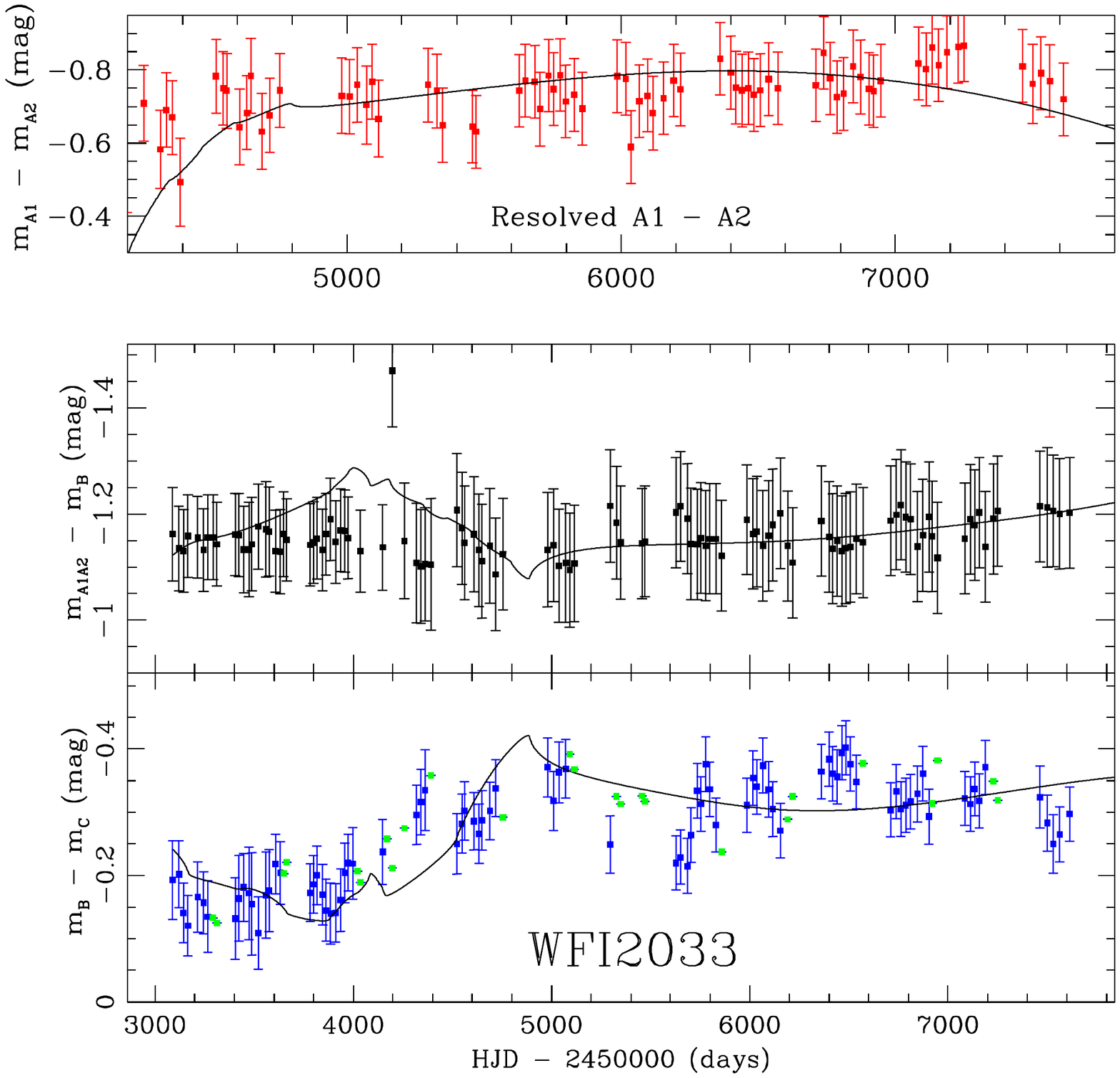}{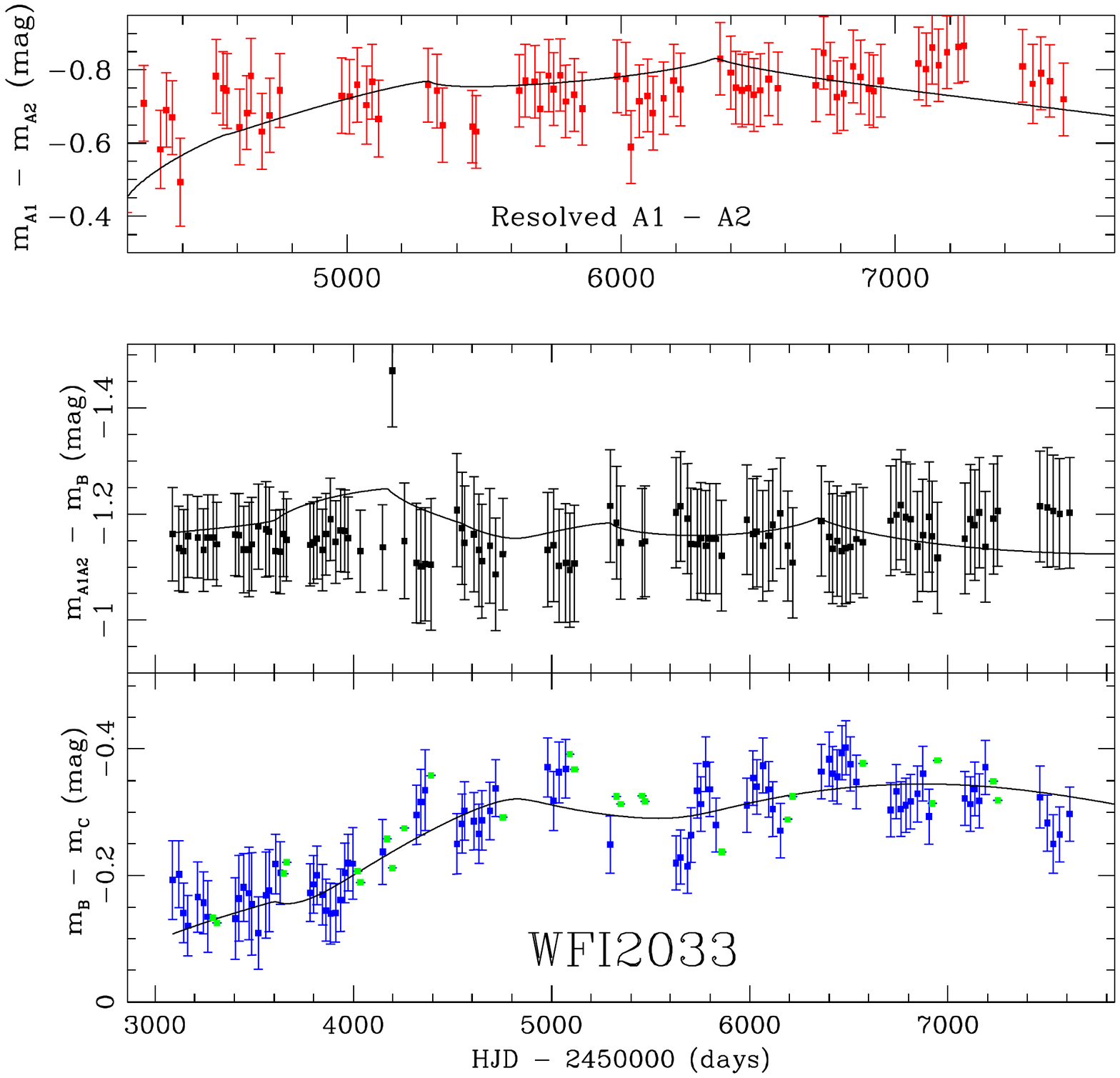}
\caption{Examples of good fits from the Monte Carlo microlensing simulation to the time-delay corrected difference light curves of WFI2033. 
The fits were produced by the $f_{M/L}=0.8$~(left panel) and 
$f_{M/L}=0.4$~(right panel) models. Points labeled in green were not used in the analysis, as they required extrapolation of $>10$~days into the inter-season gaps
 when shifting the light curves by the time delays.  Note the $\ga0.2$~mags of extrinsic variability evident in the image B-C difference light curve.} 
 \label{fig:bestfits}
 \end{figure}

Prior to each Monte Carlo trial, 
we convolved each set of magnification patterns with a Gaussian surface brightness profile at a range of trial source sizes 
\begin{equation}
14.5 \le \log({\hat r_s} \langle M_*/M_{\sun}\rangle^{-1/2}/{\rm cm}) \le 18.0,
\end{equation} 
where $\hat{r_s}$ is the radius of the accretion disk scaled by $\langle M_* \rangle^{1/2}$, the mean mass of a lens galaxy star.   
Although we used a Gaussian profile, the exact choice 
of photometric emission model is unimportant, since \citet{mortonson05} showed that microlensing statistics are largely a function of the half-light radius of the emitting region, not the 
exact properties of the emission profile. In a given trial, a convolved pattern is run past a model point source on a random trajectory and at 
a random transverse speed 10~km/s ~$\le ({\hat v}_e \, \langle M_*/M_{\sun} \rangle^{-1/2}) \le 10^6 $~km/s. 
Changes in magnification
with time are logged at the epochs of the observed data and a running comparison with the light curves is made.  The quality of the fit is tallied in real time using a $\chi^2$ statistic, and, to save 
computational time, fits with
$\chi^2/N_{dof} > 3.0$ are discarded since they will not contribute significantly to the Bayesian integrals in our post-run analysis.  
During the curve fitting process, we allowed for 0.07 and 0.03~magnitudes of systematic error in the photometry of images A1 \& A2 and B \& C, respectively.  We
also allowed for 0.5~magnitudes of uncertainty in the intrinsic flux ratios between the lensed images, since both the continuum and the 
broad line region are affected by microlensing in this lens \citep{Motta2017} and to allow for the influence of substructure in the lens.
We attempted $10^7$ fits per set of magnification patterns for a grand total of $N_{\rm trials} = 4 \times 10^9$ trials, requiring approximately two weeks of run time on the US Naval
Academy High Performance Computing (HPC) cluster.
In Figure~\ref{fig:bestfits}, we display two of the best fits from our Monte Carlo analysis to the time-delay corrected difference light curves of WFI2033.  Consistent with the findings of
 \citet{Giannini2017}, we easily see $\sim0.2$~mags of microlensing variability in the difference light curves from image C. The microlensing in images A1, A2 and B
is less pronounced, with $\lesssim0.1$~mags of extrinsic variability over the 13 seasons of monitoring.

\begin{figure}
\plottwo{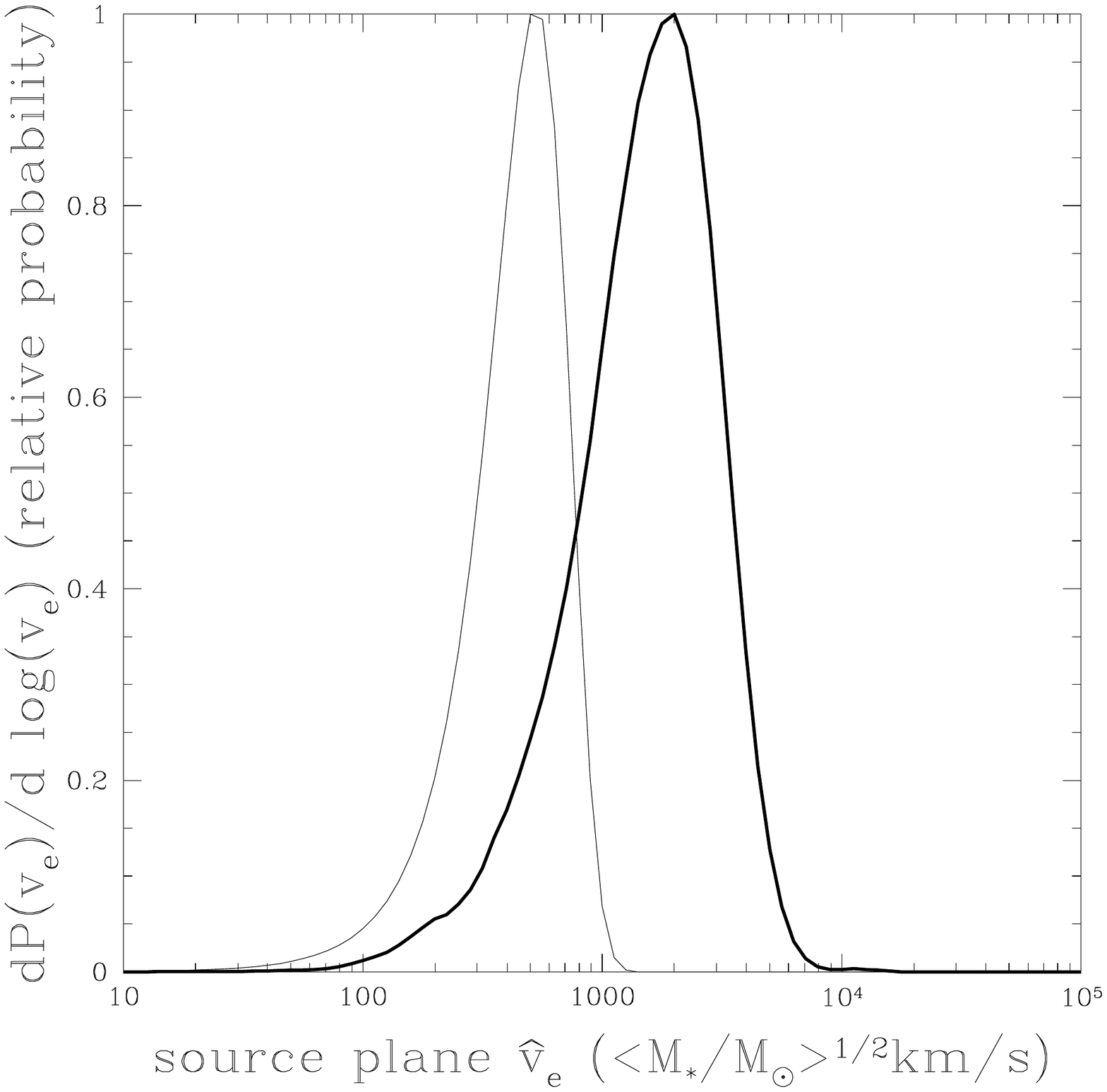}{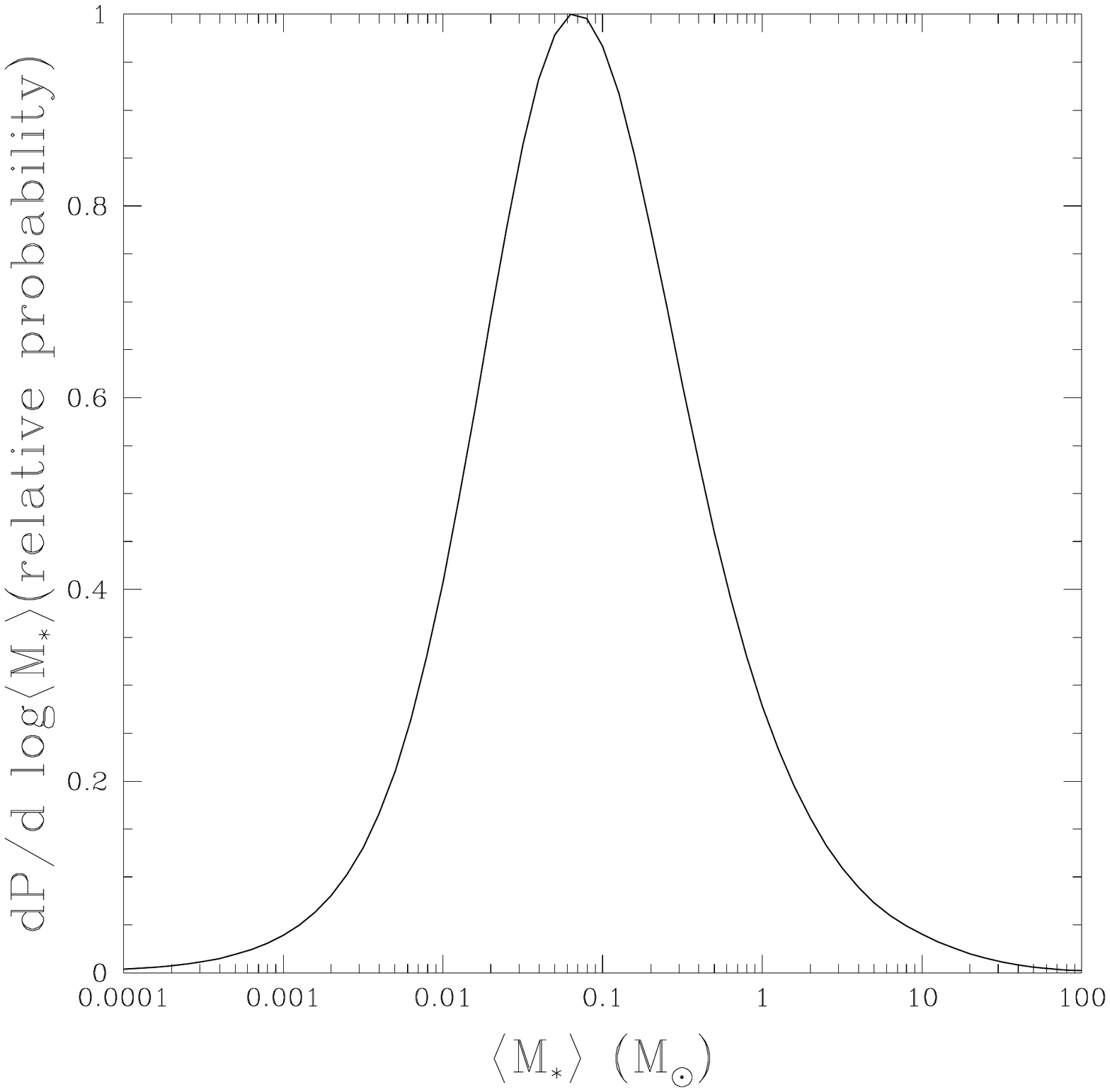}
\caption{Left: Probability density for the effective
source velocity $\hat{v}_{e}$ for WFI2033.  The heavy curve is the scaled effective velocity distribution
in Einstein units with median $\hat{v}_{e} = 1.5\times10^3\,\langle M_*/M_{\sun} \rangle^{1/2}\kms$. The thinner curve indicates the prior probability
distribution for the true source velocity $v_e$, which we construct
using the method described in \citet{Kochanek2004}.  Right: The convolution of the prior on $v_e$ with the probability density for $\hat{v}_{e}$ yields
the probability density for 
$\langle M_* \rangle$, which we use to convert the source size measurement into physical units, independent of a prior on $\langle M_* \rangle$.
Our estimate for the median lens galaxy stellar mass is $\langle M_* / M_{\sun} \rangle = 0.08^{+0.36}_{-0.05}$.
}\label{fig:ve}
\end{figure}

\subsection{Bayesian Analysis of Monte Carlo Results}

Using Bayes' theorem, the probability of the parameters given the data $D$ is 
\begin{equation}
P(\xi_{p},\xi_{t} \vert D) \propto P(D \vert \xi_{p},\xi_{t})P(\xi_{p})P(\xi_{t}), 
\end{equation} 
where $\xi_{p}$ is the collection of physical variables, $\xi_{t}$ is the collection of trajectory variables and $P(\xi_{t})$ and $P(\xi_{p})$ are the prior probabilities for the trajectory and physical variables, respectively. The physical variables are parameters of the local magnification tensors (mean convergence $\kappa$ and mean shear $\gamma$), the local properties of the stars (surface density of stars $\kappa_{\star}$, mass of the average microlens $\langle M_* \rangle$),
the scale radius of the source $r_{s}$, and the effective velocity of the source $\vec{v}_{e}$.  The probability distribution for any variable of interest can be 
obtained by marginalizing over the other variables of the simulation.  For example, to find the probability density for the scale radius $r_s$, 
\begin{equation}
   P( r_s | D) \propto \int P(D| \vec{p}, r_s ) P(\vec{p}) P(r_s) d\vec{p}
\end{equation}
where $P(D|\vec{p}, r_s)$ is the probability of fitting the data
in a particular trial, $P(\vec{p})$ sets the priors on the microlensing
variables $\xi_{p}$ \& $\xi_{t}$, and $P(r_s)$
is the (uniform) prior on the scale radius.  The total probability is
then normalized so that $\int P(r_s|D)d r_s=1$. 

\begin{figure}
\plottwo{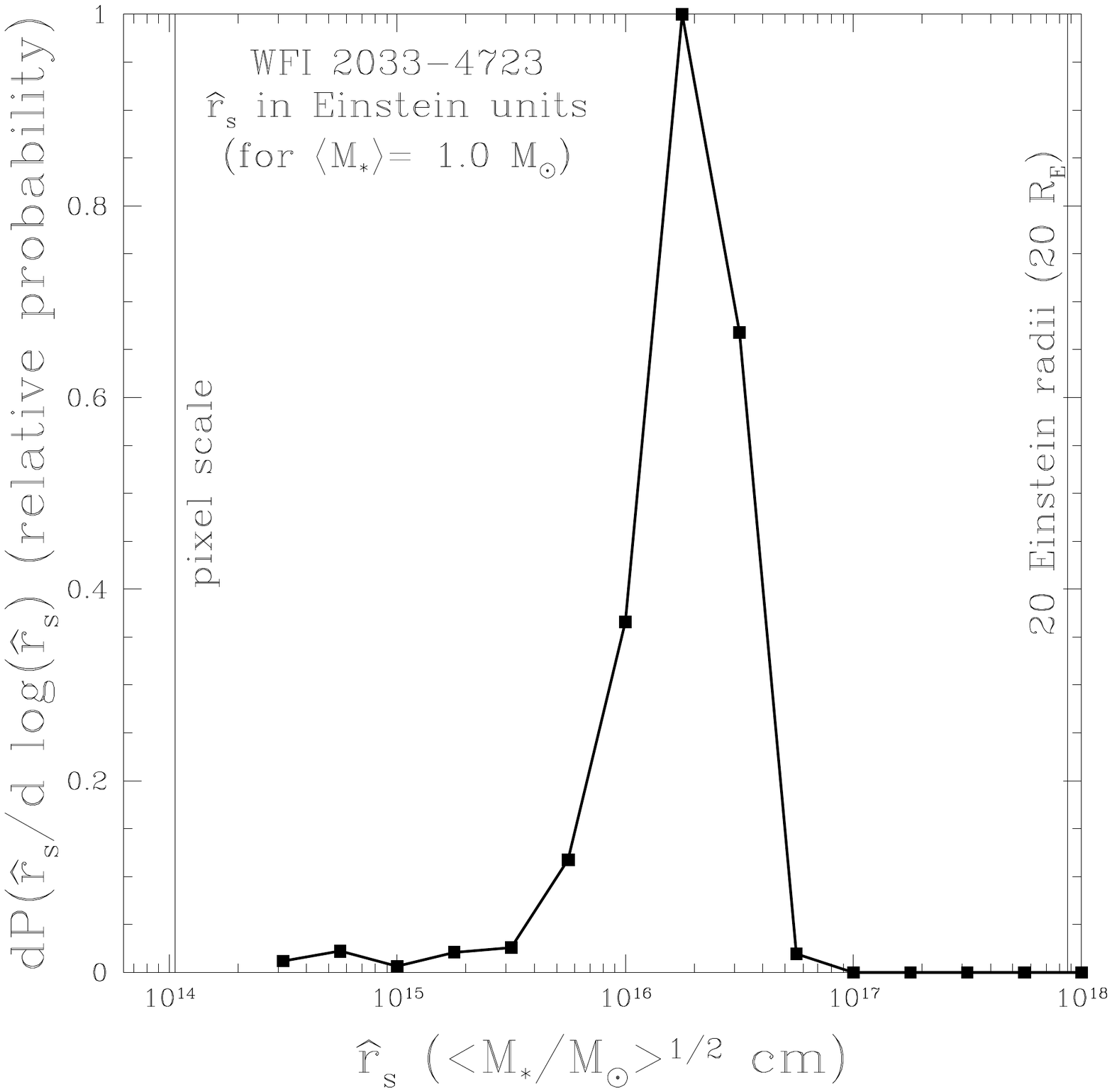}{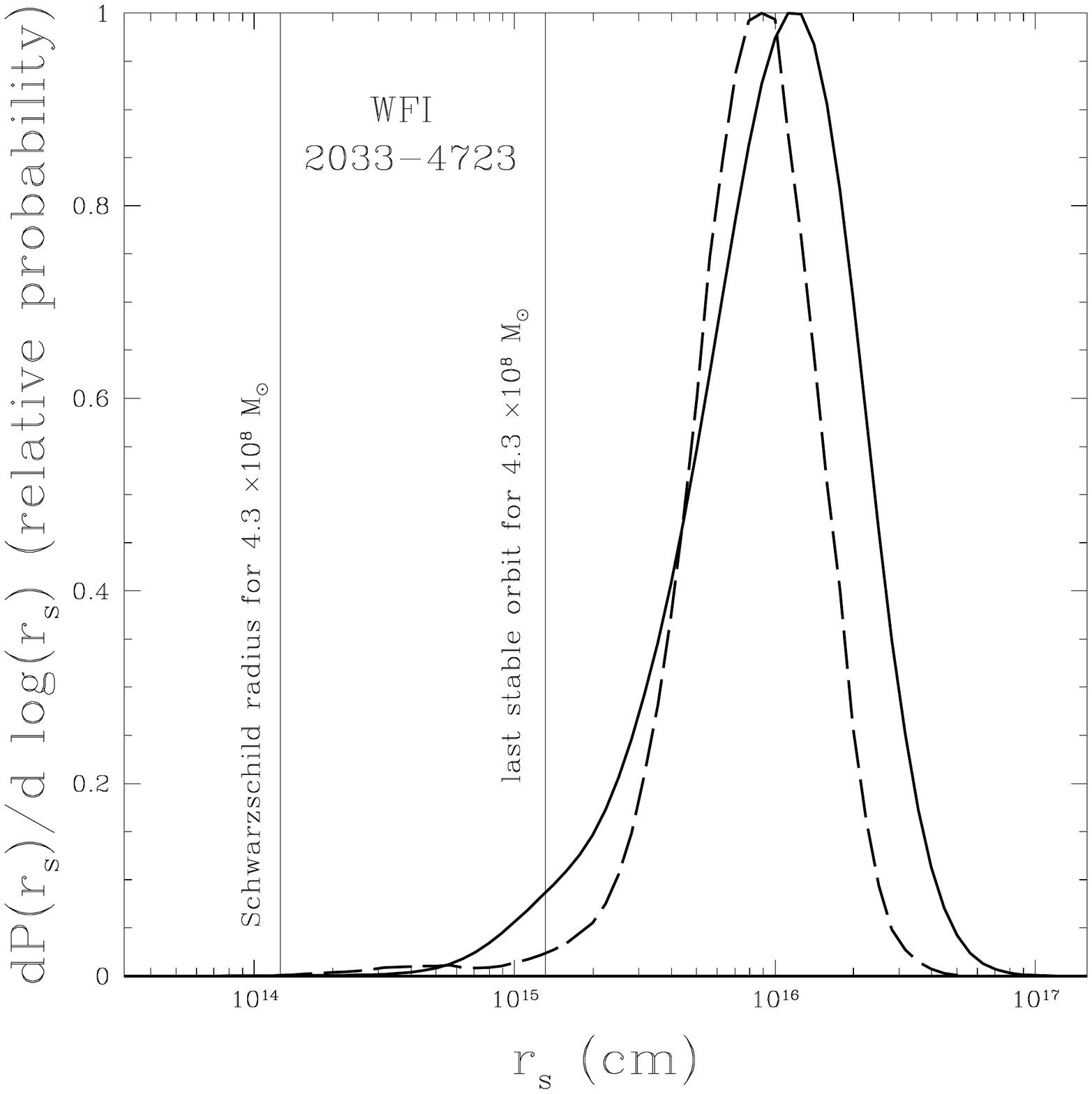}
\caption{Left: Probability density for the Gaussian scale radius in Einstein units ${\hat r}_s = r_s \langle M_*/M_{\sun} \rangle^{-1/2}$.
Right: Relative probability density for the thin disk scale radius $r_{s}$ in physical units for WFI2033 at $\lambda_{rest}=2632$~\AA.  
The solid line represents the probability density arising from the 
microlensing simulations using the prior on the true effective velocity $v_e$, while the dotted line shows the result of imposing 
a uniform prior on the mean microlens mass of $0.1 < \langle M_*/\msun \rangle < 1.0$. 
}\label{fig:rs}
\end{figure}

We initially do the calculation in ``Einstein units", where all lengths depend upon a factor of the unknown mean stellar mass,
$\langle M_*/M_{\sun} \rangle^{1/2}$.
For example, in Figure~\ref{fig:rs}, we display 
the probability density for the accretion disk scale radius ${\hat r}_s = r_s \langle M_*/M_{\sun} \rangle^{-1/2}$ in Einstein Units in which the plotted values assume $M_* = 1.0 M_{\sun}$.  
This degeneracy
can be broken, however, by examining Figure~\ref{fig:ve}, where we display the probability density for the scaled effective velocity
${\hat v}_e = v_e \langle M_*/M_{\sun} \rangle^{-1/2}$ (Einstein Units) from the Monte Carlo simulation.  We also display a model in physical units for the expected transverse 
velocity $dP(v_{e})/d\log(v_e)$ which serves as the statistical prior on $v_e$.  Since ${\hat v}_e = v_e \langle M_*/M_{\sun} \rangle^{-1/2}$, we convolve the prior on $v_e$ with 
the probability density  for ${\hat v}_e$
to produce a probability density for $\langle M_* \rangle$.  The probability density for $\langle M_* \rangle$ is then used to convert all scaled lengths (e.g. ${\hat r}_s$) into
true, physical units (e.g. $r_s$) by convolving $dP(M_*)/d\log(M_*)$ with the quantity of interest.

We construct the prior on the transverse velocity, $dP(v_{e})/d\log(v_e)$, using the method described in \citet{Kochanek2004}. 
For the peculiar velocity components of both lens and source, we make redshift-based estimates
from the models of \citet{mosquerakochanek11}.
We estimate the velocity dispersion of the lens galaxy from its Einstein radius, 
assuming the galaxy is a singular isothermal sphere with relaxed dynamics, which
\citet{treu04} and \citet{bolton08} show is a good approximation. 

We display the probability density for the scale radius in physical units $r_s$ in Figure~\ref{fig:rs}.  In this plot, we also show a probability density for $r_s$ obtained by assuming
a uniform prior on the median lens galaxy stellar mass of $0.1 \le \langle M_*/M_{\sun} \rangle \le 1.0$.  A brief inspection of the plot reveals that the results without a prior on
the microlens mass $\langle M_* \rangle$ are robustly consistent with the results using the uniform mass prior.
As a final step we must correct the scale radius for the disk's inclination $i$ by 
multiplying by $(\cos i)^{-1/2}$, which is necessary because we have assumed a 
face-on disk in our simulations and microlensing amplitudes depend on the 
projected area of a source rather than the shape.  We adopt the measurement made without the mass prior
$\log \left\{ (r_s/{\rm cm})[\cos(i)/0.5]^{1/2} \right\}= 15.86^{+0.25}_{-0.27}$ at $\lambda_{rest} = 2481{\rm \AA}$, where $i$ is the inclination angle.

\section{RESULTS \& DISCUSSION}\label{sec:results}

\begin{figure}
\begin{center}
\vskip -144pt
\plotone{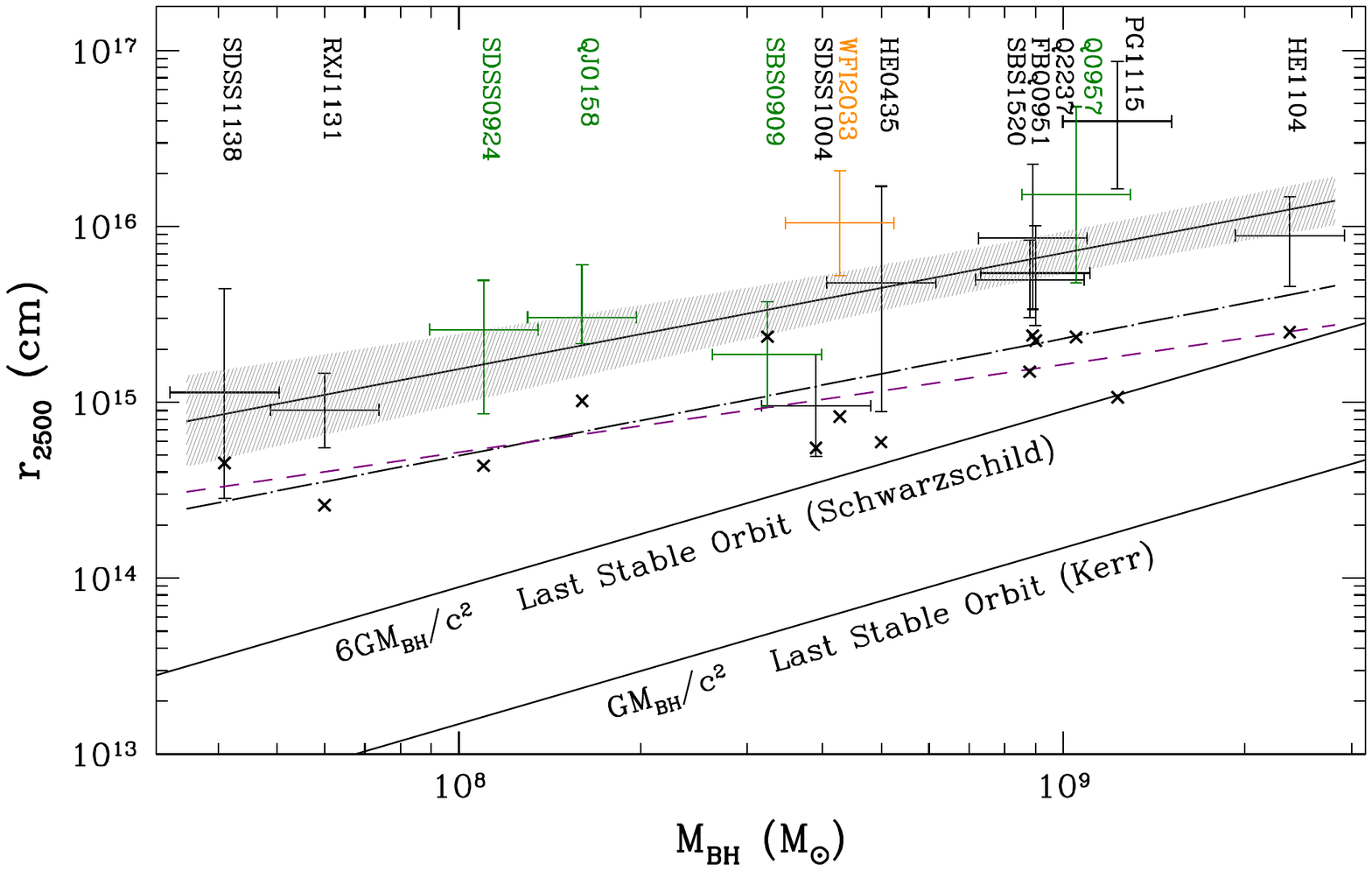}
\vskip -90pt
\caption{Accretion disk size versus supermassive black hole mass relation 
(thick solid line) and data
from \citet{morgan10} with new measurements of $r_{s}$ for Q~0957+561 \citep{hainline12}, QJ~0158--4723 \citep{morgan12}, 
SBS~0909+532 \citep{hainline13}  and SDSS~0924+0219 \citep{macleod15} (plotted in green) and WFI2033 (plotted in orange), all scaled to 
2500\,\AA\ and corrected to $60\degr$ inclination. The dash-shaded region indicates the $1\sigma$ boundaries from uncertainties 
in the slope and intercept.
The black dot-dashed
line shows the scale radius as a function of central black hole mass predicted
by theoretical thin disk models (for $L/L_{E} = 1/3$ and $\eta = 0.1$), while the small diagonal crosses indicate
the thin disk size predicted by the magnification-corrected luminosity of the 
different quasars.  The dashed purple line is a fit to the luminosity-based thin disk size estimates (diagonal crosses). 
The microlensing
source size for WFI2033 is larger than the luminosity-constrained thin disk size and the theoretical thin disk size based on black hole mass, 
similar to the findings of \citet{pooley07}, \citet{morgan10} and 
\citet{mediavilla11}.}\label{fig:accrdisk}
\end{center}
\end{figure}

In Figure~\ref{fig:accrdisk}, we plot the size of the accretion disk in WFI2033 on the Accretion Disk Size - Black Hole Mass Relation \citep{morgan10} assuming an inclination angle 
$i=60^{\circ}$, 
where we have corrected the scale radius $r_s$ at the wavelength corresponding to the center of the rest-frame $R$-band, $\lambda_{rest}=\lambda_{eff, R} / (1 + z_s) = 2481$~\AA , to 
$r_{2500}$, the scale radius at $\lambda_{rest} = 2500$~\AA, assuming the $r_s \propto \lambda^{4/3}$ scaling of \citet{shakura73} thin disk theory. We assume 
$\langle \cos(i) \rangle = 0.5$ or $\langle i \rangle =60^{\circ}$, the expectation value for the inclination of a randomly oriented disk. 
 For the black hole mass ($M_{BH}$),
we use the result from \citet{sluse12}, who used the \ion{Mg}{2} emission line to find $\log(M_{BH}/M_{\sun}) = 8.63 \pm 0.35$.  \citet{Motta2017} also estimated $M_{BH}$
in this system using Keplerian dynamics, but their method yielded very large uncertainties with $M_{BH-Motta17}=1.2^{+3.1}_{-0.8} \times 10^8 \, {\rm M_{\sun}}$. The \citet{morgan10} relation 
was derived using $M_{BH}$ estimates
based on emission line widths so the emission line width-based measurement from \citet{sluse12} is a better choice. 

The expectation value for the scale radius at $\lambda_{rest}=2681{\rm \AA}$  without the prior on the mass of microlenses 
is $\log \left\{ (r_s/{\rm cm})[\cos(i)/0.5]^{1/2} \right\}= 15.86^{+0.25}_{-0.27}$. 
This is fully consistent with the results of \citet{Motta2017}, who estimate a scale radius $r_s \approx 1.6^{+0.5}_{-0.4} \times 10^{16}$~cm at $\lambda_{rest}=1310$~\AA \, using single-epoch
spectroscopy, which, when scaled to
2481~\AA \, assuming $r_s \propto \lambda^{4/3}$ is $\log(r_s/{\rm cm}) = 15.8^{+0.2}_{-0.1}$.  The \citet{Motta2017} result is strongly dependent upon priors, especially the assumption of a median microlens mass $\langle M_*/M_{\sun}\rangle = 0.3$, nevertheless, the independence of the techniques provides robust support for our result.   \citet{blackburne11} also estimated the scale radius
of the accretion disk in WFI2033 using single epoch, multi-wavelength photometry, but their results are, by their own admission, anomalous, as they also predict with highest likelihood an
accretion disk with an inverted (increasing toward the outer edge) temperature profile. Like \cite{Motta2017}, our size measurement is significantly smaller than the \citet{blackburne11}
estimate.

In Figure~\ref{fig:accrdisk}, we made several updates in addition to the new WFI2033 measurement. Measurements of the accretion disk scale radius using the microlensing
variability technique of Q~0957+561 \citep{hainline12},  SBS~0909+532 \citep{hainline13} were added, and updates to the QJ~0158--4723 \citep{morgan12}, 
 and SDSS~0924+0219 \citep{macleod15} measurements were also included.  With changes to 2 out of the 11 existing points and the addition of 3 new measurements, the updated 
 Accretion Disk Size - Black Hole Mass Relation \citep{morgan10} is
 \begin{equation}
 \log(r_{2500}/{\rm cm})=(15.85\pm0.12) + (0.66\pm0.15)\log(M_{BH}/10^9{\rm M_{\sun}}).
 \end{equation}
This is consistent with the original fit from \citet{morgan10}, $\log(R_{2500}/$cm$)=(15.78\pm0.12) + (0.80\pm0.17)\log(M_{BH}/10^9$M$_{\sun})$, 
and the shallower slope brings the relation into excellent agreement with the expectation from thin disk theory ($r_s \propto M_{BH}^{2/3}$). 

There are now 14 systems in which the accretion disk size has been measured using the microlensing variability \citep[e.g.][]{Kochanek2004} technique.  With the exception of SBS~0909+532,
in which the luminosity-based size estimate is marginally larger than the microlensing-based size measurement, microlensing-based 
size measurements are consistently larger than the luminosity-based thin disk size estimates by an average of $\sim0.55$~dex.  
The SBS~0909+532 luminosity-based thin disk size estimate is somewhat suspect, however, 
since \citet{sluse12} and \citet{lehar00} found very different photometric fits
for the lens galaxy in this system, leading to significant uncertainty in the magnification and, consequently, the intrinsic luminosity.  
Very recent continuum emission region reverberation mapping studies in local, lower luminosity AGN \citep[e.g.][]{fasnaugh18,edelson17,mchardy16} 
have revealed similar discrepancies between observed accretion disk size measurements and the predictions of thin disk theory. 
In \citet{morgan10}, we proposed that real accretion disks lack the necessary surface brightness to produce their observed luminosity from the smaller area of a simple thin disk model, and 
we remain confident in that conclusion.  We were nevertheless intrigued to find that the slope of $r_s$ vs $M_{BH}$ is remarkably consistent with the 
predictions of thin disk theory ($r_s \propto M_{BH}^{2/3}$), so it is the intercept in the accretion disk-size black hole mass relation that is inconsistent with thin disk theory, rather than 
the slope.  In \citet{morgan10}, we suggested that the most promising explanation for the discrepancy is that accretion disks may have shallower temperature slopes  
than that predicted by thin disk theory $T(r) \propto r^{3/4}$, and this hypothesis remains fully viable.  We are hopeful that our ongoing lensed quasar monitoring campaign in the infrared 
($J-$, $H-$ and $K-$band),
corresponding to optical emission in the rest frame, will allow for measurements of accretion disk temperature profiles in the near future.

\acknowledgements

This material is based upon work supported by the National Science Foundation under
grant AST-1614018 to C.W.M. and grants AST-1515876 \& AST-1814440 to C.S.K. 
$COSMOGRAIL$ is made possible thanks to the continuous work of all observers and technical staff obtaining the monitoring observations, 
in particular at the Swiss Leonhard Euler telescope at La Silla Observatory, which is supported by the Swiss National Science Foundation.
V. Bonvin and F. Courbin are also supported by the Swiss National Science Foundation (SNSF).  
This work is partly based on 
observations obtained with the Small and Moderate Aperture Research Telescope 
System (SMARTS) 1.3m, which is operated by the SMARTS Consortium. Thanks to V. Motta for helpful discussions about
black hole masses and to R. Witt for many hours of help with the USNA High Performance Computing Cluster.

\pagebreak

\def\hm{\hphantom{-}}
\begin{deluxetable}{ccccccc}
\tabletypesize{\scriptsize}
\tablecaption{WFI2033--4723 Light Curves - SMARTS}
\tablewidth{0pt}
\tablehead{ HJD
                &\multicolumn{1}{c}{A1} &\multicolumn{1}{c}{A2} 
                &\multicolumn{1}{c}{B} &\multicolumn{1}{c}{C}
                &\multicolumn{1}{c}{$\chi^2/N_{dof}$}
                &\multicolumn{1}{c}{Source} 
              }
\startdata
$3082.897$ & $2.747\pm 0.012$ & $3.396\pm 0.021$ & $3.425\pm 0.012$ & $3.640\pm 0.015$ & $1.6$ & SMARTS \\ 
$3112.857$ & $2.769\pm 0.017$ & $3.411\pm 0.029$ & $3.554\pm 0.014$ & $3.638\pm 0.017$ & $1.7$ & SMARTS \\ 
$3138.866$ & $2.835\pm 0.017$ & $3.460\pm 0.030$ & $3.610\pm 0.016$ & $3.708\pm 0.019$ & $3.0$ & SMARTS \\ 
$3146.900$ & $2.848\pm 0.015$ & $3.448\pm 0.025$ & $3.606\pm 0.013$ & $3.718\pm 0.015$ & $5.7$ & SMARTS \\ 
$3154.840$ & $2.870\pm 0.015$ & $3.511\pm 0.025$ & $3.635\pm 0.013$ & $3.670\pm 0.015$ & $2.9$ & SMARTS \\ 
$3175.825$ & $2.937\pm 0.011$ & $3.502\pm 0.016$ & $3.584\pm 0.011$ & $3.751\pm 0.013$ & $6.0$ & SMARTS \\ 
$3184.777$ & $2.948\pm 0.014$ & $3.519\pm 0.023$ & $3.601\pm 0.013$ & $3.748\pm 0.016$ & $1.8$ & SMARTS \\ 
$3211.740$ & $2.921\pm 0.014$ & $3.572\pm 0.025$ & $3.568\pm 0.013$ & $3.737\pm 0.016$ & $1.7$ & SMARTS \\ 
$3282.634$ & $2.996\pm 0.014$ & $3.444\pm 0.020$ & $3.544\pm 0.011$ & $3.780\pm 0.014$ & $3.0$ & SMARTS \\ 
$3295.616$ & $2.892\pm 0.013$ & $3.566\pm 0.023$ & $3.544\pm 0.012$ & $3.794\pm 0.015$ & $4.2$ & SMARTS \\ 
$3298.566$ & $2.931\pm 0.015$ & $3.548\pm 0.025$ & $3.564\pm 0.014$ & $3.787\pm 0.017$ & $1.4$ & SMARTS \\ 
$3310.551$ & $2.881\pm 0.013$ & $3.578\pm 0.024$ & $3.605\pm 0.013$ & $3.789\pm 0.015$ & $1.5$ & SMARTS \\ 
$3320.551$ & $3.081\pm 0.021$ & $3.273\pm 0.024$ & $3.602\pm 0.015$ & $3.831\pm 0.019$ & $4.0$ & SMARTS \\ 
$3592.759$ & $2.966\pm 0.016$ & $3.607\pm 0.028$ & $3.528\pm 0.014$ & $3.890\pm 0.019$ & $1.8$ & SMARTS \\ 
$3625.668$ & $2.881\pm 0.014$ & $3.531\pm 0.024$ & $3.463\pm 0.015$ & $3.854\pm 0.022$ & $1.5$ & SMARTS \\ 
$3651.552$ & $2.838\pm 0.017$ & $3.475\pm 0.030$ & $3.423\pm 0.013$ & $3.824\pm 0.021$ & $2.0$ & SMARTS \\ 
$3661.571$ & $2.893\pm 0.031$ & $3.339\pm 0.045$ & $3.445\pm 0.024$ & $3.875\pm 0.038$ & $1.0$ & SMARTS \\ 
$3665.553$ & $2.904\pm 0.017$ & $3.363\pm 0.024$ & $3.440\pm 0.013$ & $3.801\pm 0.018$ & $2.1$ & SMARTS \\ 
$3675.522$ & $2.895\pm 0.018$ & $3.315\pm 0.025$ & $3.411\pm 0.015$ & $3.800\pm 0.020$ & $2.2$ & SMARTS \\ 
$3826.891$ & $2.762\pm 0.018$ & $3.483\pm 0.034$ & $3.436\pm 0.016$ & $3.681\pm 0.020$ & $1.0$ & SMARTS \\ 
$3832.855$ & $2.742\pm 0.024$ & $3.453\pm 0.045$ & $3.553\pm 0.020$ & $3.723\pm 0.026$ & $4.0$ & SMARTS \\ 
$3852.870$ & $2.740\pm 0.015$ & $3.495\pm 0.028$ & $3.524\pm 0.015$ & $3.697\pm 0.018$ & $1.9$ & SMARTS \\ 
$3863.788$ & $2.736\pm 0.016$ & $3.471\pm 0.030$ & $3.570\pm 0.018$ & $3.679\pm 0.020$ & $4.7$ & SMARTS \\ 
$3886.859$ & $2.819\pm 0.012$ & $3.486\pm 0.020$ & $3.599\pm 0.014$ & $3.666\pm 0.015$ & $1.8$ & SMARTS \\ 
$3937.732$ & $2.942\pm 0.016$ & $3.578\pm 0.028$ & $3.517\pm 0.014$ & $3.772\pm 0.018$ & $1.6$ & SMARTS \\ 
$3994.640$ & $2.938\pm 0.019$ & $3.281\pm 0.025$ & $3.499\pm 0.014$ & $3.785\pm 0.019$ & $2.3$ & SMARTS \\ 
$4021.575$ & $2.932\pm 0.018$ & $3.325\pm 0.027$ & $3.598\pm 0.016$ & $3.757\pm 0.020$ & $1.5$ & SMARTS \\ 
$4042.528$ & $2.889\pm 0.027$ & $3.478\pm 0.046$ & $3.532\pm 0.025$ & $3.738\pm 0.032$ & $0.9$ & SMARTS \\ 
$4050.555$ & $2.921\pm 0.016$ & $3.510\pm 0.026$ & $3.494\pm 0.015$ & $3.765\pm 0.018$ & $4.4$ & SMARTS \\ 
$4064.511$ & $2.963\pm 0.041$ & $3.462\pm 0.064$ & $3.397\pm 0.036$ & $3.728\pm 0.050$ & $0.6$ & SMARTS \\ 
$4207.876$ & $2.773\pm 0.019$ & $3.472\pm 0.036$ & $3.540\pm 0.016$ & $3.791\pm 0.022$ & $3.5$ & SMARTS \\ 
$4224.867$ & $2.832\pm 0.023$ & $3.400\pm 0.038$ & $3.456\pm 0.020$ & $3.763\pm 0.028$ & $1.8$ & SMARTS \\ 
$4234.918$ & $2.811\pm 0.018$ & $3.471\pm 0.031$ & $3.374\pm 0.014$ & $3.695\pm 0.020$ & $0.9$ & SMARTS \\ 
$4243.898$ & $2.818\pm 0.015$ & $3.450\pm 0.026$ & $3.385\pm 0.013$ & $3.713\pm 0.018$ & $1.3$ & SMARTS \\ 
$4293.829$ & $2.703\pm 0.015$ & $3.413\pm 0.028$ & $3.399\pm 0.013$ & $3.649\pm 0.017$ & $1.3$ & SMARTS \\ 
$4345.711$ & $2.758\pm 0.021$ & $3.381\pm 0.036$ & $3.353\pm 0.016$ & $3.700\pm 0.023$ & $2.1$ & SMARTS \\ 
$4363.622$ & $2.796\pm 0.038$ & $3.255\pm 0.057$ & $3.304\pm 0.031$ & $3.703\pm 0.048$ & $0.3$ & SMARTS \\ 
$4367.667$ & $2.839\pm 0.030$ & $3.259\pm 0.042$ & $3.345\pm 0.024$ & $3.675\pm 0.035$ & $0.9$ & SMARTS \\ 
$4371.624$ & $2.768\pm 0.024$ & $3.395\pm 0.043$ & $3.357\pm 0.020$ & $3.685\pm 0.029$ & $0.8$ & SMARTS \\ 
$4378.583$ & $2.687\pm 0.014$ & $3.449\pm 0.026$ & $3.343\pm 0.012$ & $3.680\pm 0.017$ & $1.7$ & SMARTS \\ 
$4387.542$ & $2.690\pm 0.014$ & $3.414\pm 0.026$ & $3.343\pm 0.013$ & $3.659\pm 0.017$ & $1.6$ & SMARTS \\ 
$4390.519$ & $2.706\pm 0.012$ & $3.407\pm 0.021$ & $3.324\pm 0.012$ & $3.648\pm 0.016$ & $1.8$ & SMARTS \\ 
$4394.534$ & $2.696\pm 0.023$ & $3.404\pm 0.043$ & $3.311\pm 0.021$ & $3.719\pm 0.031$ & $0.9$ & SMARTS \\ 
$4397.544$ & $2.725\pm 0.020$ & $3.276\pm 0.032$ & $3.271\pm 0.017$ & $3.697\pm 0.025$ & $1.0$ & SMARTS \\ 
$4407.520$ & $2.693\pm 0.025$ & $3.384\pm 0.046$ & $3.257\pm 0.018$ & $3.661\pm 0.028$ & $1.0$ & SMARTS \\ 
$4427.512$ & $2.739\pm 0.037$ & $3.233\pm 0.057$ & $3.322\pm 0.032$ & $3.680\pm 0.047$ & $0.7$ & SMARTS \\ 
$4550.907$ & $2.589\pm 0.021$ & $3.286\pm 0.039$ & $3.349\pm 0.020$ & $3.630\pm 0.028$ & $3.8$ & SMARTS \\ 
$4557.869$ & $2.587\pm 0.013$ & $3.304\pm 0.024$ & $3.321\pm 0.012$ & $3.623\pm 0.016$ & $1.4$ & SMARTS \\ 
$4564.904$ & $2.519\pm 0.011$ & $3.373\pm 0.021$ & $3.339\pm 0.011$ & $3.618\pm 0.014$ & $3.0$ & SMARTS \\ 
$4571.811$ & $2.554\pm 0.015$ & $3.369\pm 0.031$ & $3.346\pm 0.013$ & $3.569\pm 0.017$ & $1.2$ & SMARTS \\ 
$4588.916$ & $2.637\pm 0.015$ & $3.331\pm 0.027$ & $3.373\pm 0.013$ & $3.602\pm 0.017$ & $2.5$ & SMARTS \\ 
$4589.840$ & $2.603\pm 0.014$ & $3.354\pm 0.025$ & $3.385\pm 0.012$ & $3.570\pm 0.015$ & $4.4$ & SMARTS \\ 
$4596.807$ & $2.637\pm 0.012$ & $3.379\pm 0.022$ & $3.393\pm 0.011$ & $3.587\pm 0.014$ & $1.7$ & SMARTS \\ 
$4633.820$ & $2.704\pm 0.021$ & $3.350\pm 0.037$ & $3.374\pm 0.017$ & $3.703\pm 0.026$ & $2.6$ & SMARTS \\ 
$4653.812$ & $2.706\pm 0.020$ & $3.350\pm 0.036$ & $3.379\pm 0.016$ & $3.658\pm 0.023$ & $0.9$ & SMARTS \\ 
$4660.793$ & $2.660\pm 0.015$ & $3.403\pm 0.029$ & $3.392\pm 0.013$ & $3.677\pm 0.019$ & $1.6$ & SMARTS \\ 
$4678.731$ & $2.744\pm 0.013$ & $3.391\pm 0.022$ & $3.376\pm 0.011$ & $3.681\pm 0.015$ & $4.4$ & SMARTS \\ 
$4684.695$ & $2.698\pm 0.012$ & $3.482\pm 0.022$ & $3.378\pm 0.011$ & $3.630\pm 0.015$ & $2.6$ & SMARTS \\ 
$4716.731$ & $2.667\pm 0.031$ & $3.375\pm 0.058$ & $3.363\pm 0.026$ & $3.690\pm 0.038$ & $0.6$ & SMARTS \\ 
$4724.697$ & $2.725\pm 0.020$ & $3.348\pm 0.035$ & $3.328\pm 0.019$ & $3.676\pm 0.026$ & $1.8$ & SMARTS \\ 
$4732.649$ & $2.729\pm 0.025$ & $3.331\pm 0.042$ & $3.324\pm 0.019$ & $3.684\pm 0.027$ & $1.2$ & SMARTS \\ 
$4747.635$ & $2.729\pm 0.017$ & $3.432\pm 0.032$ & $3.388\pm 0.014$ & $3.691\pm 0.019$ & $1.2$ & SMARTS \\ 
$4754.586$ & $2.765\pm 0.022$ & $3.314\pm 0.036$ & $3.379\pm 0.018$ & $3.659\pm 0.024$ & $0.5$ & SMARTS \\ 
$4758.587$ & $2.703\pm 0.012$ & $3.408\pm 0.022$ & $3.359\pm 0.011$ & $3.671\pm 0.014$ & $3.7$ & SMARTS \\ 
$4783.521$ & $2.705\pm 0.016$ & $3.410\pm 0.029$ & $3.349\pm 0.016$ & $3.651\pm 0.021$ & $1.0$ & SMARTS \\ 
$4790.543$ & $2.682\pm 0.012$ & $3.456\pm 0.023$ & $3.339\pm 0.011$ & $3.690\pm 0.015$ & $3.4$ & SMARTS \\ 
$4797.550$ & $2.686\pm 0.014$ & $3.422\pm 0.026$ & $3.366\pm 0.012$ & $3.686\pm 0.015$ & $2.2$ & SMARTS \\ 
$5009.684$ & $2.463\pm 0.014$ & $3.193\pm 0.026$ & $3.135\pm 0.011$ & $3.551\pm 0.016$ & $4.5$ & SMARTS \\ 
$5021.762$ & $2.461\pm 0.022$ & $3.191\pm 0.043$ & $2.993\pm 0.020$ & $3.407\pm 0.031$ & $1.1$ & SMARTS \\ 
$5038.667$ & $2.430\pm 0.014$ & $3.130\pm 0.025$ & $3.071\pm 0.010$ & $3.549\pm 0.016$ & $4.8$ & SMARTS \\ 
$5043.671$ & $2.421\pm 0.021$ & $3.196\pm 0.042$ & $3.023\pm 0.022$ & $3.478\pm 0.035$ & $0.5$ & SMARTS \\ 
$5053.723$ & $2.417\pm 0.016$ & $3.153\pm 0.030$ & $2.987\pm 0.012$ & $3.480\pm 0.020$ & $1.7$ & SMARTS \\ 
$5072.652$ & $2.393\pm 0.011$ & $3.151\pm 0.021$ & $3.020\pm 0.010$ & $3.433\pm 0.015$ & $1.8$ & SMARTS \\ 
$5106.592$ & $2.370\pm 0.019$ & $3.074\pm 0.035$ & $3.038\pm 0.015$ & $3.415\pm 0.022$ & $1.2$ & SMARTS \\ 
$5127.553$ & $2.372\pm 0.012$ & $3.140\pm 0.022$ & $3.092\pm 0.011$ & $3.379\pm 0.014$ & $2.6$ & SMARTS \\ 
$5150.518$ & $2.423\pm 0.017$ & $3.090\pm 0.031$ & $3.130\pm 0.013$ & $3.421\pm 0.018$ & $2.1$ & SMARTS \\ 
$5326.806$ & $2.420\pm 0.014$ & $3.212\pm 0.028$ & $3.232\pm 0.012$ & $3.443\pm 0.016$ & $1.3$ & SMARTS \\ 
$5335.836$ & $2.449\pm 0.013$ & $3.183\pm 0.023$ & $3.236\pm 0.012$ & $3.483\pm 0.015$ & $2.1$ & SMARTS \\ 
$5353.802$ & $2.473\pm 0.010$ & $3.237\pm 0.019$ & $3.271\pm 0.011$ & $3.445\pm 0.012$ & $2.9$ & SMARTS \\ 
$5372.779$ & $2.542\pm 0.017$ & $3.229\pm 0.032$ & $3.254\pm 0.017$ & $3.514\pm 0.023$ & $0.5$ & SMARTS \\ 
$5379.848$ & $2.533\pm 0.016$ & $3.272\pm 0.030$ & $3.280\pm 0.014$ & $3.503\pm 0.018$ & $1.0$ & SMARTS \\ 
$5388.792$ & $2.627\pm 0.017$ & $3.193\pm 0.027$ & $3.152\pm 0.011$ & $3.601\pm 0.018$ & $7.8$ & SMARTS \\ 
$6407.908$ & $2.534\pm 0.021$ & $3.311\pm 0.041$ & $3.103\pm 0.017$ & $3.546\pm 0.026$ & $2.8$ & SMARTS \\ 
$6418.889$ & $2.511\pm 0.012$ & $3.257\pm 0.021$ & $3.136\pm 0.011$ & $3.635\pm 0.015$ & $5.1$ & SMARTS \\ 
$6431.908$ & $2.498\pm 0.014$ & $3.225\pm 0.027$ & $3.123\pm 0.011$ & $3.659\pm 0.018$ & $8.7$ & SMARTS \\ 
$6436.834$ & $2.496\pm 0.017$ & $3.298\pm 0.035$ & $3.177\pm 0.016$ & $3.627\pm 0.025$ & $1.8$ & SMARTS \\ 
$6458.851$ & $2.546\pm 0.013$ & $3.336\pm 0.025$ & $3.180\pm 0.011$ & $3.559\pm 0.015$ & $4.1$ & SMARTS \\ 
$6464.838$ & $2.534\pm 0.017$ & $3.373\pm 0.035$ & $3.053\pm 0.014$ & $3.578\pm 0.023$ & $11.8$ & SMARTS \\ 
$6488.762$ & $2.620\pm 0.017$ & $3.276\pm 0.030$ & $3.230\pm 0.013$ & $3.578\pm 0.018$ & $2.4$ & SMARTS \\ 
$6492.691$ & $2.587\pm 0.017$ & $3.282\pm 0.030$ & $3.186\pm 0.014$ & $3.598\pm 0.022$ & $5.0$ & SMARTS \\ 
$6508.687$ & $2.589\pm 0.045$ & $3.311\pm 0.088$ & $3.142\pm 0.036$ & $3.606\pm 0.060$ & $0.9$ & SMARTS \\ 
$6760.883$ & $2.546\pm 0.016$ & $3.251\pm 0.030$ & $3.212\pm 0.014$ & $3.589\pm 0.020$ & $3.3$ & SMARTS \\ 
$6825.834$ & $2.594\pm 0.024$ & $3.401\pm 0.049$ & $3.289\pm 0.019$ & $3.632\pm 0.029$ & $1.2$ & SMARTS \\ 
$6857.805$ & $2.589\pm 0.017$ & $3.367\pm 0.033$ & $3.179\pm 0.012$ & $3.665\pm 0.020$ & $3.2$ & SMARTS \\ 
$6944.600$ & $2.546\pm 0.016$ & $3.309\pm 0.031$ & $3.131\pm 0.011$ & $3.596\pm 0.017$ & $4.9$ & SMARTS \\ 
$7141.898$ & $2.541\pm 0.014$ & $3.320\pm 0.026$ & $3.153\pm 0.012$ & $3.622\pm 0.018$ & $2.5$ & SMARTS \\ 
$7150.873$ & $2.607\pm 0.033$ & $3.209\pm 0.057$ & $3.149\pm 0.027$ & $3.683\pm 0.048$ & $2.1$ & SMARTS \\ 
$7253.669$ & $2.420\pm 0.013$ & $3.237\pm 0.026$ & $3.245\pm 0.012$ & $3.503\pm 0.016$ & $2.3$ & SMARTS \\ 
$7255.638$ & $2.406\pm 0.023$ & $3.228\pm 0.047$ & $3.274\pm 0.021$ & $3.503\pm 0.030$ & $0.7$ & SMARTS \\ 
$7269.616$ & $2.339\pm 0.014$ & $3.304\pm 0.032$ & $3.267\pm 0.013$ & $3.577\pm 0.019$ & $3.7$ & SMARTS \\ 
$7278.583$ & $2.458\pm 0.014$ & $3.275\pm 0.028$ & $3.266\pm 0.013$ & $3.519\pm 0.018$ & $2.0$ & SMARTS \\ 
$7340.526$ & $2.459\pm 0.012$ & $3.374\pm 0.025$ & $3.225\pm 0.012$ & $3.556\pm 0.016$ & $1.7$ & SMARTS \\ 
$7344.525$ & $2.509\pm 0.023$ & $3.241\pm 0.044$ & $3.222\pm 0.024$ & $3.630\pm 0.036$ & $0.9$ & SMARTS \\ 
$7598.716$ & $2.611\pm 0.021$ & $3.338\pm 0.040$ & $3.341\pm 0.020$ & $3.605\pm 0.027$ & $0.6$ & SMARTS \\ 
$7603.680$ & $2.619\pm 0.015$ & $3.355\pm 0.027$ & $3.364\pm 0.012$ & $3.607\pm 0.017$ & $1.5$ & SMARTS \\ 
$7605.790$ & $2.586\pm 0.015$ & $3.347\pm 0.028$ & $3.319\pm 0.013$ & $3.633\pm 0.017$ & $4.1$ & SMARTS \\ 
$7608.765$ & $2.620\pm 0.017$ & $3.323\pm 0.031$ & $3.289\pm 0.013$ & $3.636\pm 0.019$ & $1.1$ & SMARTS \\ 
$7613.738$ & $2.583\pm 0.018$ & $3.406\pm 0.037$ & $3.317\pm 0.016$ & $3.614\pm 0.022$ & $1.1$ & SMARTS \\ 
$7652.639$ & $2.665\pm 0.020$ & $3.274\pm 0.033$ & $3.211\pm 0.015$ & $3.462\pm 0.020$ & $7.7$ & SMARTS \\ 
$7661.601$ & $2.634\pm 0.019$ & $3.272\pm 0.032$ & $3.320\pm 0.014$ & $3.547\pm 0.020$ & $2.3$ & SMARTS \\ 
$7695.556$ & $2.662\pm 0.017$ & $3.344\pm 0.031$ & $3.231\pm 0.013$ & $3.587\pm 0.020$ & $3.1$ & SMARTS \\ 
$7702.583$ & $2.610\pm 0.019$ & $3.373\pm 0.038$ & $3.302\pm 0.018$ & $3.547\pm 0.023$ & $2.4$ & SMARTS \\ 
$7856.868$ & $2.866\pm 0.032$ & $3.130\pm 0.038$ & $3.531\pm 0.026$ & $3.762\pm 0.035$ & $1.0$ & SMARTS \\ 
$7867.882$ & $2.676\pm 0.015$ & $3.471\pm 0.029$ & $3.536\pm 0.015$ & $3.691\pm 0.018$ & $2.5$ & SMARTS \\ 
$7872.884$ & $2.687\pm 0.014$ & $3.478\pm 0.027$ & $3.452\pm 0.014$ & $3.736\pm 0.019$ & $2.1$ & SMARTS \\ 
$7877.840$ & $2.700\pm 0.017$ & $3.516\pm 0.034$ & $3.497\pm 0.017$ & $3.701\pm 0.022$ & $1.6$ & SMARTS \\ 
$7894.808$ & $2.717\pm 0.020$ & $3.462\pm 0.038$ & $3.511\pm 0.018$ & $3.723\pm 0.023$ & $3.4$ & SMARTS \\ 
$7904.766$ & $2.697\pm 0.016$ & $3.493\pm 0.031$ & $3.520\pm 0.016$ & $3.714\pm 0.020$ & $4.4$ & SMARTS \\ 
$7914.785$ & $2.725\pm 0.024$ & $3.372\pm 0.041$ & $3.499\pm 0.027$ & $3.756\pm 0.035$ & $0.7$ & SMARTS \\
\enddata
\tablecomments{HJD is the Heliocentric Julian Day --2450000 days.
The goodness of fit of the image, $\chi^2/N_{dof}$, is used to rescale the
formal uncertainties by a factor of $(\chi^2/N_{dof})^{1/2}$.  The Image A1-C
columns give the magnitudes of the quasar images relative to the
comparison stars.}
\label{tab:lcsmarts}
\end{deluxetable}

\def\hm{\hphantom{-}}
\begin{deluxetable}{ccccccc}
\tabletypesize{\scriptsize}
\tablecaption{WFI2033--4723 Light Curves - EULER}
\tablewidth{0pt}
\tablehead{ HJD
                &\multicolumn{1}{c}{A1} &\multicolumn{1}{c}{A2} 
                &\multicolumn{1}{c}{B} &\multicolumn{1}{c}{C}
                &\multicolumn{1}{c}{$\chi^2/N_{dof}$}
                &\multicolumn{1}{c}{Source} 
              }
\startdata
$5485.659$ & $2.628\pm 0.008$ & $3.285\pm 0.013$ & $3.348\pm 0.007$ & $3.597\pm 0.009$ & $2.4$ & EULER \\ 
$5488.637$ & $2.644\pm 0.008$ & $3.281\pm 0.013$ & $3.341\pm 0.007$ & $3.579\pm 0.008$ & $3.8$ & EULER \\ 
$5503.592$ & $2.654\pm 0.006$ & $3.294\pm 0.009$ & $3.368\pm 0.005$ & $3.615\pm 0.006$ & $10.5$ & EULER \\ 
$5506.571$ & $2.662\pm 0.007$ & $3.293\pm 0.010$ & $3.378\pm 0.006$ & $3.629\pm 0.007$ & $6.3$ & EULER \\ 
$5655.905$ & $2.764\pm 0.002$ & $3.518\pm 0.004$ & $3.505\pm 0.003$ & $3.684\pm 0.004$ & $2.8$ & EULER \\ 
$5656.910$ & $2.742\pm 0.007$ & $3.564\pm 0.013$ & $3.499\pm 0.007$ & $3.697\pm 0.008$ & $6.1$ & EULER \\ 
$5667.908$ & $2.768\pm 0.001$ & $3.506\pm 0.002$ & $3.589\pm 0.002$ & $3.732\pm 0.002$ & $6.6$ & EULER \\ 
$5674.897$ & $2.792\pm 0.008$ & $3.534\pm 0.014$ & $3.569\pm 0.008$ & $3.729\pm 0.009$ & $2.2$ & EULER \\ 
$5678.879$ & $2.772\pm 0.007$ & $3.528\pm 0.012$ & $3.575\pm 0.006$ & $3.733\pm 0.007$ & $7.8$ & EULER \\ 
$5682.896$ & $2.773\pm 0.007$ & $3.541\pm 0.012$ & $3.587\pm 0.007$ & $3.731\pm 0.008$ & $5.7$ & EULER \\ 
$5686.849$ & $2.791\pm 0.007$ & $3.518\pm 0.012$ & $3.583\pm 0.007$ & $3.742\pm 0.008$ & $3.4$ & EULER \\ 
$5694.851$ & $2.760\pm 0.007$ & $3.587\pm 0.012$ & $3.602\pm 0.007$ & $3.744\pm 0.008$ & $11.0$ & EULER \\ 
$5712.864$ & $2.832\pm 0.006$ & $3.536\pm 0.009$ & $3.524\pm 0.006$ & $3.829\pm 0.007$ & $8.7$ & EULER \\ 
$5723.776$ & $2.860\pm 0.008$ & $3.565\pm 0.013$ & $3.495\pm 0.007$ & $3.786\pm 0.009$ & $2.6$ & EULER \\ 
$5725.771$ & $2.828\pm 0.002$ & $3.616\pm 0.005$ & $3.500\pm 0.004$ & $3.784\pm 0.005$ & $3.7$ & EULER \\ 
$5739.722$ & $2.861\pm 0.009$ & $3.554\pm 0.016$ & $3.431\pm 0.007$ & $3.799\pm 0.009$ & $2.5$ & EULER \\ 
$5762.714$ & $2.758\pm 0.008$ & $3.549\pm 0.016$ & $3.437\pm 0.008$ & $3.792\pm 0.011$ & $3.1$ & EULER \\ 
$5766.821$ & $2.746\pm 0.009$ & $3.510\pm 0.016$ & $3.387\pm 0.006$ & $3.865\pm 0.010$ & $3.4$ & EULER \\ 
$5770.741$ & $2.723\pm 0.007$ & $3.534\pm 0.013$ & $3.430\pm 0.006$ & $3.807\pm 0.009$ & $3.7$ & EULER \\ 
$5775.621$ & $2.785\pm 0.016$ & $3.381\pm 0.028$ & $3.449\pm 0.013$ & $3.791\pm 0.018$ & $1.1$ & EULER \\ 
$5779.607$ & $2.710\pm 0.007$ & $3.515\pm 0.013$ & $3.431\pm 0.006$ & $3.783\pm 0.008$ & $3.4$ & EULER \\ 
$5783.712$ & $2.725\pm 0.007$ & $3.448\pm 0.013$ & $3.414\pm 0.008$ & $3.824\pm 0.011$ & $2.8$ & EULER \\ 
$5794.685$ & $2.704\pm 0.006$ & $3.473\pm 0.011$ & $3.430\pm 0.006$ & $3.805\pm 0.008$ & $4.5$ & EULER \\ 
$5804.569$ & $2.712\pm 0.008$ & $3.503\pm 0.015$ & $3.501\pm 0.007$ & $3.716\pm 0.009$ & $2.9$ & EULER \\ 
$5807.536$ & $2.711\pm 0.001$ & $3.500\pm 0.002$ & $3.501\pm 0.002$ & $3.720\pm 0.002$ & $4.5$ & EULER \\ 
$5815.552$ & $2.702\pm 0.008$ & $3.500\pm 0.015$ & $3.528\pm 0.009$ & $3.808\pm 0.011$ & $4.9$ & EULER \\ 
$5818.636$ & $2.732\pm 0.007$ & $3.454\pm 0.011$ & $3.493\pm 0.007$ & $3.803\pm 0.009$ & $5.3$ & EULER \\ 
$5820.656$ & $2.741\pm 0.007$ & $3.452\pm 0.011$ & $3.500\pm 0.007$ & $3.803\pm 0.009$ & $5.0$ & EULER \\ 
$5824.700$ & $2.755\pm 0.009$ & $3.432\pm 0.015$ & $3.507\pm 0.007$ & $3.798\pm 0.009$ & $4.5$ & EULER \\ 
$5827.557$ & $2.715\pm 0.008$ & $3.438\pm 0.013$ & $3.575\pm 0.007$ & $3.835\pm 0.010$ & $8.9$ & EULER \\ 
$5831.537$ & $2.748\pm 0.005$ & $3.464\pm 0.008$ & $3.545\pm 0.006$ & $3.773\pm 0.007$ & $16.3$ & EULER \\ 
$5839.610$ & $2.783\pm 0.011$ & $3.467\pm 0.020$ & $3.545\pm 0.009$ & $3.851\pm 0.013$ & $3.4$ & EULER \\ 
$5842.521$ & $2.775\pm 0.009$ & $3.536\pm 0.016$ & $3.583\pm 0.009$ & $3.805\pm 0.012$ & $2.7$ & EULER \\ 
$5854.529$ & $2.813\pm 0.008$ & $3.544\pm 0.014$ & $3.585\pm 0.007$ & $3.794\pm 0.009$ & $4.1$ & EULER \\ 
$5857.517$ & $2.830\pm 0.007$ & $3.580\pm 0.013$ & $3.576\pm 0.006$ & $3.763\pm 0.008$ & $3.2$ & EULER \\ 
$5865.500$ & $2.849\pm 0.018$ & $3.579\pm 0.034$ & $3.608\pm 0.019$ & $3.799\pm 0.024$ & $0.9$ & EULER \\ 
$5865.512$ & $2.852\pm 0.009$ & $3.534\pm 0.015$ & $3.577\pm 0.008$ & $3.785\pm 0.010$ & $2.8$ & EULER \\ 
$5869.535$ & $2.819\pm 0.009$ & $3.617\pm 0.017$ & $3.583\pm 0.008$ & $3.777\pm 0.010$ & $2.5$ & EULER \\ 
$5873.574$ & $2.884\pm 0.017$ & $3.509\pm 0.030$ & $3.563\pm 0.013$ & $3.847\pm 0.019$ & $1.3$ & EULER \\ 
$5887.557$ & $2.916\pm 0.007$ & $3.587\pm 0.013$ & $3.542\pm 0.007$ & $3.816\pm 0.008$ & $3.2$ & EULER \\ 
$5896.532$ & $2.907\pm 0.010$ & $3.653\pm 0.019$ & $3.548\pm 0.009$ & $3.810\pm 0.011$ & $1.5$ & EULER \\ 
$5897.530$ & $2.912\pm 0.011$ & $3.600\pm 0.020$ & $3.525\pm 0.009$ & $3.841\pm 0.012$ & $2.4$ & EULER \\ 
$6011.897$ & $2.783\pm 0.001$ & $3.567\pm 0.001$ & $3.540\pm 0.001$ & $3.775\pm 0.001$ & $5.0$ & EULER \\ 
$6015.909$ & $2.759\pm 0.009$ & $3.585\pm 0.019$ & $3.602\pm 0.009$ & $3.758\pm 0.011$ & $7.1$ & EULER \\ 
$6017.894$ & $2.778\pm 0.007$ & $3.558\pm 0.013$ & $3.561\pm 0.007$ & $3.757\pm 0.008$ & $8.2$ & EULER \\ 
$6018.905$ & $2.808\pm 0.008$ & $3.554\pm 0.015$ & $3.550\pm 0.007$ & $3.763\pm 0.009$ & $2.3$ & EULER \\ 
$6023.906$ & $2.796\pm 0.009$ & $3.601\pm 0.018$ & $3.597\pm 0.009$ & $3.780\pm 0.011$ & $4.9$ & EULER \\ 
$6028.900$ & $2.822\pm 0.007$ & $3.581\pm 0.012$ & $3.589\pm 0.007$ & $3.826\pm 0.009$ & $4.8$ & EULER \\ 
$6029.911$ & $2.828\pm 0.009$ & $3.587\pm 0.016$ & $3.585\pm 0.008$ & $3.812\pm 0.010$ & $2.4$ & EULER \\ 
$6047.869$ & $2.811\pm 0.008$ & $3.596\pm 0.015$ & $3.588\pm 0.007$ & $3.829\pm 0.009$ & $7.3$ & EULER \\ 
$6050.863$ & $2.821\pm 0.008$ & $3.559\pm 0.014$ & $3.581\pm 0.007$ & $3.865\pm 0.009$ & $6.6$ & EULER \\ 
$6058.888$ & $2.810\pm 0.008$ & $3.617\pm 0.014$ & $3.565\pm 0.007$ & $3.914\pm 0.011$ & $4.9$ & EULER \\ 
$6070.935$ & $2.891\pm 0.007$ & $3.480\pm 0.010$ & $3.496\pm 0.007$ & $3.897\pm 0.009$ & $9.5$ & EULER \\ 
$6092.942$ & $2.864\pm 0.011$ & $3.522\pm 0.019$ & $3.516\pm 0.008$ & $3.878\pm 0.012$ & $3.6$ & EULER \\ 
$6102.712$ & $2.797\pm 0.007$ & $3.577\pm 0.012$ & $3.622\pm 0.007$ & $3.891\pm 0.008$ & $8.5$ & EULER \\ 
$6109.823$ & $2.854\pm 0.008$ & $3.515\pm 0.013$ & $3.575\pm 0.008$ & $3.929\pm 0.010$ & $7.4$ & EULER \\ 
$6125.707$ & $2.830\pm 0.008$ & $3.577\pm 0.015$ & $3.649\pm 0.007$ & $3.902\pm 0.010$ & $6.0$ & EULER \\ 
$6129.706$ & $2.857\pm 0.008$ & $3.579\pm 0.015$ & $3.632\pm 0.009$ & $3.894\pm 0.012$ & $5.8$ & EULER \\ 
$6138.590$ & $2.882\pm 0.011$ & $3.600\pm 0.020$ & $3.673\pm 0.010$ & $3.865\pm 0.013$ & $3.3$ & EULER \\ 
$6151.591$ & $2.899\pm 0.009$ & $3.581\pm 0.017$ & $3.630\pm 0.008$ & $3.893\pm 0.010$ & $2.7$ & EULER \\ 
$6185.503$ & $2.854\pm 0.012$ & $3.592\pm 0.021$ & $3.569\pm 0.012$ & $3.942\pm 0.016$ & $3.0$ & EULER \\ 
$6185.515$ & $2.855\pm 0.007$ & $3.616\pm 0.011$ & $3.567\pm 0.007$ & $3.930\pm 0.009$ & $7.4$ & EULER \\ 
$6190.528$ & $2.867\pm 0.007$ & $3.591\pm 0.013$ & $3.548\pm 0.007$ & $3.940\pm 0.009$ & $6.8$ & EULER \\ 
$6194.537$ & $2.864\pm 0.013$ & $3.623\pm 0.025$ & $3.533\pm 0.011$ & $3.919\pm 0.018$ & $1.3$ & EULER \\ 
$6195.693$ & $2.901\pm 0.007$ & $3.562\pm 0.013$ & $3.466\pm 0.007$ & $3.914\pm 0.011$ & $3.7$ & EULER \\ 
$6217.615$ & $2.860\pm 0.008$ & $3.545\pm 0.014$ & $3.405\pm 0.006$ & $3.925\pm 0.009$ & $6.2$ & EULER \\ 
$6221.532$ & $2.845\pm 0.009$ & $3.566\pm 0.015$ & $3.441\pm 0.007$ & $3.931\pm 0.010$ & $3.0$ & EULER \\ 
$6224.554$ & $2.819\pm 0.011$ & $3.594\pm 0.022$ & $3.439\pm 0.010$ & $3.915\pm 0.015$ & $1.4$ & EULER \\ 
$6225.544$ & $2.821\pm 0.002$ & $3.583\pm 0.005$ & $3.473\pm 0.004$ & $3.895\pm 0.006$ & $1.4$ & EULER \\ 
$6232.501$ & $2.831\pm 0.002$ & $3.635\pm 0.005$ & $3.400\pm 0.004$ & $3.825\pm 0.005$ & $0.8$ & EULER \\ 
$6232.513$ & $2.794\pm 0.010$ & $3.588\pm 0.020$ & $3.456\pm 0.008$ & $3.869\pm 0.012$ & $1.3$ & EULER \\ 
$6236.552$ & $2.816\pm 0.007$ & $3.513\pm 0.012$ & $3.392\pm 0.006$ & $3.891\pm 0.008$ & $5.7$ & EULER \\ 
$6248.524$ & $2.814\pm 0.008$ & $3.511\pm 0.013$ & $3.339\pm 0.006$ & $3.837\pm 0.009$ & $6.1$ & EULER \\ 
$6251.522$ & $2.773\pm 0.007$ & $3.548\pm 0.013$ & $3.348\pm 0.006$ & $3.824\pm 0.009$ & $5.2$ & EULER \\ 
$6255.521$ & $2.795\pm 0.005$ & $3.552\pm 0.009$ & $3.351\pm 0.006$ & $3.811\pm 0.010$ & $1.5$ & EULER \\ 
$6387.892$ & $2.479\pm 0.006$ & $3.290\pm 0.011$ & $3.221\pm 0.006$ & $3.618\pm 0.007$ & $7.7$ & EULER \\ 
$6391.909$ & $2.507\pm 0.007$ & $3.293\pm 0.014$ & $3.223\pm 0.007$ & $3.622\pm 0.009$ & $2.9$ & EULER \\ 
$6396.909$ & $2.470\pm 0.006$ & $3.336\pm 0.012$ & $3.229\pm 0.006$ & $3.597\pm 0.008$ & $6.5$ & EULER \\ 
$6401.916$ & $2.483\pm 0.007$ & $3.317\pm 0.013$ & $3.244\pm 0.007$ & $3.619\pm 0.009$ & $6.7$ & EULER \\ 
$6405.897$ & $2.481\pm 0.008$ & $3.338\pm 0.016$ & $3.201\pm 0.008$ & $3.597\pm 0.011$ & $1.5$ & EULER \\ 
$6426.900$ & $2.491\pm 0.010$ & $3.308\pm 0.018$ & $3.261\pm 0.009$ & $3.582\pm 0.014$ & $1.8$ & EULER \\ 
$6435.872$ & $2.502\pm 0.006$ & $3.288\pm 0.010$ & $3.276\pm 0.006$ & $3.643\pm 0.008$ & $6.9$ & EULER \\ 
$6443.779$ & $2.504\pm 0.013$ & $3.299\pm 0.027$ & $3.295\pm 0.013$ & $3.580\pm 0.018$ & $1.3$ & EULER \\ 
$6447.755$ & $2.507\pm 0.006$ & $3.321\pm 0.012$ & $3.284\pm 0.006$ & $3.574\pm 0.007$ & $4.3$ & EULER \\ 
$6451.826$ & $2.546\pm 0.006$ & $3.282\pm 0.011$ & $3.284\pm 0.006$ & $3.618\pm 0.008$ & $3.0$ & EULER \\ 
$6455.899$ & $2.563\pm 0.001$ & $3.311\pm 0.003$ & $3.253\pm 0.002$ & $3.624\pm 0.003$ & $1.9$ & EULER \\ 
$6460.692$ & $2.555\pm 0.009$ & $3.353\pm 0.017$ & $3.296\pm 0.008$ & $3.595\pm 0.010$ & $2.9$ & EULER \\ 
$6468.728$ & $2.553\pm 0.008$ & $3.347\pm 0.016$ & $3.317\pm 0.008$ & $3.600\pm 0.010$ & $2.6$ & EULER \\ 
$6472.831$ & $2.558\pm 0.007$ & $3.313\pm 0.011$ & $3.252\pm 0.006$ & $3.645\pm 0.008$ & $4.6$ & EULER \\ 
$6476.840$ & $2.565\pm 0.007$ & $3.305\pm 0.013$ & $3.274\pm 0.006$ & $3.644\pm 0.008$ & $3.7$ & EULER \\ 
$6487.905$ & $2.579\pm 0.006$ & $3.289\pm 0.010$ & $3.242\pm 0.005$ & $3.571\pm 0.006$ & $5.5$ & EULER \\ 
$6491.597$ & $2.555\pm 0.008$ & $3.320\pm 0.016$ & $3.288\pm 0.007$ & $3.603\pm 0.009$ & $2.3$ & EULER \\ 
$6507.754$ & $2.583\pm 0.006$ & $3.325\pm 0.009$ & $3.252\pm 0.005$ & $3.629\pm 0.006$ & $4.7$ & EULER \\ 
$6515.817$ & $2.596\pm 0.007$ & $3.313\pm 0.012$ & $3.277\pm 0.005$ & $3.622\pm 0.007$ & $3.1$ & EULER \\ 
$6519.698$ & $2.570\pm 0.005$ & $3.276\pm 0.009$ & $3.272\pm 0.005$ & $3.705\pm 0.007$ & $15.2$ & EULER \\ 
$6523.581$ & $2.565\pm 0.008$ & $3.387\pm 0.015$ & $3.310\pm 0.007$ & $3.607\pm 0.009$ & $3.7$ & EULER \\ 
$6536.594$ & $2.559\pm 0.008$ & $3.355\pm 0.015$ & $3.290\pm 0.007$ & $3.651\pm 0.010$ & $2.6$ & EULER \\ 
$6541.644$ & $2.579\pm 0.008$ & $3.313\pm 0.013$ & $3.240\pm 0.007$ & $3.681\pm 0.010$ & $4.5$ & EULER \\ 
$6544.573$ & $2.564\pm 0.008$ & $3.345\pm 0.015$ & $3.283\pm 0.007$ & $3.674\pm 0.011$ & $2.3$ & EULER \\ 
$6548.724$ & $2.604\pm 0.008$ & $3.283\pm 0.013$ & $3.203\pm 0.006$ & $3.672\pm 0.008$ & $6.0$ & EULER \\ 
$6565.497$ & $2.567\pm 0.009$ & $3.349\pm 0.016$ & $3.253\pm 0.007$ & $3.597\pm 0.010$ & $1.8$ & EULER \\ 
$6565.509$ & $2.546\pm 0.009$ & $3.348\pm 0.018$ & $3.237\pm 0.008$ & $3.607\pm 0.011$ & $1.6$ & EULER \\ 
$6569.521$ & $2.530\pm 0.005$ & $3.353\pm 0.009$ & $3.226\pm 0.005$ & $3.647\pm 0.007$ & $5.0$ & EULER \\ 
$6572.558$ & $2.548\pm 0.005$ & $3.301\pm 0.008$ & $3.203\pm 0.005$ & $3.659\pm 0.006$ & $8.6$ & EULER \\ 
$6576.549$ & $2.523\pm 0.006$ & $3.315\pm 0.010$ & $3.221\pm 0.005$ & $3.666\pm 0.007$ & $6.3$ & EULER \\ 
$6581.604$ & $2.541\pm 0.007$ & $3.249\pm 0.013$ & $3.203\pm 0.006$ & $3.669\pm 0.009$ & $6.1$ & EULER \\ 
$6584.628$ & $2.535\pm 0.008$ & $3.281\pm 0.014$ & $3.225\pm 0.007$ & $3.615\pm 0.010$ & $3.2$ & EULER \\ 
$6599.552$ & $2.515\pm 0.006$ & $3.293\pm 0.010$ & $3.209\pm 0.005$ & $3.596\pm 0.006$ & $5.5$ & EULER \\ 
$6600.574$ & $2.518\pm 0.005$ & $3.249\pm 0.008$ & $3.198\pm 0.005$ & $3.596\pm 0.006$ & $11.9$ & EULER \\ 
$6604.538$ & $2.527\pm 0.007$ & $3.236\pm 0.013$ & $3.191\pm 0.006$ & $3.631\pm 0.008$ & $3.9$ & EULER \\ 
$6609.540$ & $2.511\pm 0.005$ & $3.262\pm 0.009$ & $3.197\pm 0.005$ & $3.590\pm 0.007$ & $8.2$ & EULER \\ 
$6612.521$ & $2.519\pm 0.008$ & $3.276\pm 0.015$ & $3.190\pm 0.007$ & $3.608\pm 0.009$ & $3.1$ & EULER \\ 
$6616.521$ & $2.524\pm 0.009$ & $3.328\pm 0.017$ & $3.216\pm 0.007$ & $3.560\pm 0.010$ & $1.8$ & EULER \\ 
$6745.906$ & $2.533\pm 0.002$ & $3.291\pm 0.003$ & $3.310\pm 0.003$ & $3.609\pm 0.004$ & $2.9$ & EULER \\ 
$6765.911$ & $2.523\pm 0.006$ & $3.388\pm 0.012$ & $3.368\pm 0.007$ & $3.555\pm 0.008$ & $8.7$ & EULER \\ 
$6775.899$ & $2.492\pm 0.006$ & $3.377\pm 0.010$ & $3.371\pm 0.006$ & $3.595\pm 0.007$ & $7.1$ & EULER \\ 
$6781.853$ & $2.540\pm 0.007$ & $3.307\pm 0.013$ & $3.378\pm 0.006$ & $3.602\pm 0.008$ & $3.5$ & EULER \\ 
$6789.900$ & $2.543\pm 0.007$ & $3.303\pm 0.013$ & $3.363\pm 0.007$ & $3.624\pm 0.009$ & $2.4$ & EULER \\ 
$6793.875$ & $2.549\pm 0.007$ & $3.327\pm 0.013$ & $3.390\pm 0.008$ & $3.636\pm 0.010$ & $4.4$ & EULER \\ 
$6797.924$ & $2.551\pm 0.009$ & $3.315\pm 0.018$ & $3.309\pm 0.010$ & $3.696\pm 0.014$ & $2.4$ & EULER \\ 
$6803.883$ & $2.581\pm 0.007$ & $3.346\pm 0.012$ & $3.334\pm 0.006$ & $3.659\pm 0.009$ & $4.2$ & EULER \\ 
$6805.832$ & $2.574\pm 0.006$ & $3.370\pm 0.011$ & $3.364\pm 0.006$ & $3.614\pm 0.007$ & $5.3$ & EULER \\ 
$6814.894$ & $2.631\pm 0.006$ & $3.331\pm 0.010$ & $3.288\pm 0.005$ & $3.655\pm 0.007$ & $6.9$ & EULER \\ 
$6818.896$ & $2.631\pm 0.009$ & $3.319\pm 0.015$ & $3.278\pm 0.009$ & $3.657\pm 0.012$ & $2.6$ & EULER \\ 
$6822.706$ & $2.647\pm 0.013$ & $3.363\pm 0.024$ & $3.353\pm 0.011$ & $3.604\pm 0.015$ & $1.3$ & EULER \\ 
$6834.749$ & $2.596\pm 0.006$ & $3.370\pm 0.010$ & $3.370\pm 0.005$ & $3.648\pm 0.007$ & $6.8$ & EULER \\ 
$6846.826$ & $2.596\pm 0.006$ & $3.331\pm 0.010$ & $3.268\pm 0.005$ & $3.677\pm 0.007$ & $8.1$ & EULER \\ 
$6874.560$ & $2.578\pm 0.009$ & $3.390\pm 0.017$ & $3.279\pm 0.007$ & $3.623\pm 0.010$ & $1.7$ & EULER \\ 
$6888.633$ & $2.540\pm 0.006$ & $3.350\pm 0.011$ & $3.261\pm 0.006$ & $3.671\pm 0.008$ & $5.0$ & EULER \\ 
$6908.528$ & $2.539\pm 0.009$ & $3.320\pm 0.018$ & $3.292\pm 0.009$ & $3.598\pm 0.012$ & $1.8$ & EULER \\ 
$6930.589$ & $2.531\pm 0.006$ & $3.301\pm 0.010$ & $3.204\pm 0.005$ & $3.646\pm 0.007$ & $5.2$ & EULER \\ 
$6937.568$ & $2.531\pm 0.008$ & $3.286\pm 0.015$ & $3.194\pm 0.007$ & $3.654\pm 0.011$ & $3.1$ & EULER \\ 
$6943.603$ & $2.547\pm 0.005$ & $3.267\pm 0.009$ & $3.187\pm 0.005$ & $3.614\pm 0.006$ & $8.0$ & EULER \\ 
$6947.525$ & $2.525\pm 0.008$ & $3.300\pm 0.014$ & $3.203\pm 0.007$ & $3.647\pm 0.010$ & $3.1$ & EULER \\ 
$6950.609$ & $2.558\pm 0.008$ & $3.274\pm 0.013$ & $3.200\pm 0.006$ & $3.601\pm 0.008$ & $3.5$ & EULER \\ 
$6954.612$ & $2.542\pm 0.005$ & $3.287\pm 0.009$ & $3.208\pm 0.005$ & $3.577\pm 0.006$ & $11.6$ & EULER \\ 
$6963.579$ & $2.515\pm 0.008$ & $3.284\pm 0.016$ & $3.192\pm 0.007$ & $3.610\pm 0.010$ & $2.7$ & EULER \\ 
$6974.526$ & $2.505\pm 0.007$ & $3.292\pm 0.014$ & $3.224\pm 0.006$ & $3.574\pm 0.008$ & $3.1$ & EULER \\ 
$6990.516$ & $2.547\pm 0.013$ & $3.272\pm 0.024$ & $3.226\pm 0.011$ & $3.549\pm 0.015$ & $1.4$ & EULER \\ 
$7113.905$ & $2.549\pm 0.008$ & $3.370\pm 0.016$ & $3.300\pm 0.007$ & $3.628\pm 0.009$ & $2.3$ & EULER \\ 
$7117.910$ & $2.562\pm 0.011$ & $3.362\pm 0.022$ & $3.334\pm 0.010$ & $3.587\pm 0.012$ & $2.2$ & EULER \\ 
$7123.909$ & $2.568\pm 0.008$ & $3.355\pm 0.016$ & $3.313\pm 0.007$ & $3.645\pm 0.010$ & $2.4$ & EULER \\ 
$7126.883$ & $2.541\pm 0.006$ & $3.376\pm 0.011$ & $3.297\pm 0.006$ & $3.639\pm 0.007$ & $5.9$ & EULER \\ 
$7138.858$ & $2.544\pm 0.007$ & $3.370\pm 0.013$ & $3.273\pm 0.006$ & $3.577\pm 0.008$ & $3.2$ & EULER \\ 
$7141.852$ & $2.531\pm 0.007$ & $3.364\pm 0.014$ & $3.269\pm 0.006$ & $3.588\pm 0.008$ & $4.6$ & EULER \\ 
$7145.844$ & $2.560\pm 0.010$ & $3.324\pm 0.019$ & $3.238\pm 0.008$ & $3.614\pm 0.011$ & $1.6$ & EULER \\ 
$7153.793$ & $2.521\pm 0.008$ & $3.354\pm 0.015$ & $3.225\pm 0.007$ & $3.603\pm 0.009$ & $4.1$ & EULER \\ 
$7157.888$ & $2.535\pm 0.005$ & $3.316\pm 0.008$ & $3.193\pm 0.005$ & $3.642\pm 0.007$ & $7.8$ & EULER \\ 
$7161.828$ & $2.507\pm 0.007$ & $3.391\pm 0.014$ & $3.224\pm 0.007$ & $3.595\pm 0.009$ & $3.6$ & EULER \\ 
$7170.829$ & $2.490\pm 0.005$ & $3.353\pm 0.009$ & $3.209\pm 0.004$ & $3.628\pm 0.006$ & $4.3$ & EULER \\ 
$7178.703$ & $2.505\pm 0.014$ & $3.247\pm 0.028$ & $3.172\pm 0.011$ & $3.578\pm 0.017$ & $1.2$ & EULER \\ 
$7186.875$ & $2.450\pm 0.006$ & $3.272\pm 0.011$ & $3.132\pm 0.005$ & $3.625\pm 0.008$ & $4.2$ & EULER \\ 
$7189.909$ & $2.460\pm 0.007$ & $3.233\pm 0.013$ & $3.108\pm 0.006$ & $3.620\pm 0.008$ & $3.8$ & EULER \\ 
$7193.878$ & $2.428\pm 0.006$ & $3.252\pm 0.010$ & $3.091\pm 0.005$ & $3.640\pm 0.007$ & $10.5$ & EULER \\ 
$7196.876$ & $2.442\pm 0.007$ & $3.226\pm 0.013$ & $3.090\pm 0.006$ & $3.652\pm 0.009$ & $5.5$ & EULER \\ 
$7200.681$ & $2.441\pm 0.007$ & $3.289\pm 0.013$ & $3.130\pm 0.006$ & $3.546\pm 0.008$ & $3.3$ & EULER \\ 
$7220.648$ & $2.414\pm 0.007$ & $3.248\pm 0.013$ & $3.204\pm 0.006$ & $3.547\pm 0.007$ & $4.4$ & EULER \\ 
$7227.605$ & $2.390\pm 0.007$ & $3.257\pm 0.014$ & $3.205\pm 0.006$ & $3.519\pm 0.008$ & $2.7$ & EULER \\ 
$7258.518$ & $2.424\pm 0.007$ & $3.286\pm 0.013$ & $3.236\pm 0.006$ & $3.457\pm 0.007$ & $4.0$ & EULER \\ 
$7263.564$ & $2.412\pm 0.008$ & $3.299\pm 0.016$ & $3.274\pm 0.009$ & $3.499\pm 0.011$ & $2.7$ & EULER \\ 
$7267.530$ & $2.429\pm 0.005$ & $3.296\pm 0.009$ & $3.272\pm 0.005$ & $3.492\pm 0.006$ & $6.5$ & EULER \\ 
$7270.537$ & $2.440\pm 0.007$ & $3.271\pm 0.014$ & $3.291\pm 0.006$ & $3.504\pm 0.008$ & $2.5$ & EULER \\ 
$7278.519$ & $2.447\pm 0.006$ & $3.315\pm 0.011$ & $3.294\pm 0.006$ & $3.501\pm 0.007$ & $4.8$ & EULER \\ 
$7293.582$ & $2.441\pm 0.006$ & $3.308\pm 0.010$ & $3.242\pm 0.006$ & $3.612\pm 0.007$ & $9.5$ & EULER \\ 
$7491.891$ & $2.518\pm 0.007$ & $3.307\pm 0.012$ & $3.306\pm 0.006$ & $3.615\pm 0.007$ & $2.4$ & EULER \\ 
$7507.870$ & $2.508\pm 0.007$ & $3.345\pm 0.014$ & $3.328\pm 0.007$ & $3.582\pm 0.008$ & $2.2$ & EULER \\ 
$7536.937$ & $2.542\pm 0.020$ & $3.304\pm 0.040$ & $3.361\pm 0.020$ & $3.643\pm 0.028$ & $0.8$ & EULER \\ 
$7557.851$ & $2.575\pm 0.006$ & $3.383\pm 0.011$ & $3.335\pm 0.006$ & $3.629\pm 0.008$ & $4.1$ & EULER \\ 
$7575.652$ & $2.577\pm 0.006$ & $3.353\pm 0.011$ & $3.361\pm 0.005$ & $3.575\pm 0.006$ & $6.3$ & EULER \\ 
$7590.677$ & $2.570\pm 0.006$ & $3.373\pm 0.011$ & $3.371\pm 0.007$ & $3.596\pm 0.009$ & $5.4$ & EULER \\ 
$7596.577$ & $2.603\pm 0.010$ & $3.332\pm 0.018$ & $3.363\pm 0.008$ & $3.593\pm 0.010$ & $2.0$ & EULER \\ 
$7599.776$ & $2.579\pm 0.007$ & $3.349\pm 0.013$ & $3.315\pm 0.006$ & $3.618\pm 0.008$ & $4.5$ & EULER \\ 
$7609.557$ & $2.600\pm 0.007$ & $3.350\pm 0.013$ & $3.340\pm 0.006$ & $3.607\pm 0.007$ & $3.9$ & EULER \\ 
$7647.656$ & $2.590\pm 0.008$ & $3.294\pm 0.015$ & $3.322\pm 0.008$ & $3.636\pm 0.010$ & $3.0$ & EULER \\ 
$7651.684$ & $2.587\pm 0.006$ & $3.314\pm 0.011$ & $3.336\pm 0.006$ & $3.618\pm 0.007$ & $7.2$ & EULER \\ 
$7671.581$ & $2.581\pm 0.002$ & $3.380\pm 0.004$ & $3.373\pm 0.003$ & $3.638\pm 0.004$ & $3.3$ & EULER \\ 
$7691.574$ & $2.630\pm 0.007$ & $3.336\pm 0.012$ & $3.329\pm 0.006$ & $3.640\pm 0.007$ & $5.1$ & EULER \\ 
$7696.536$ & $2.618\pm 0.005$ & $3.354\pm 0.008$ & $3.337\pm 0.005$ & $3.589\pm 0.006$ & $21.9$ & EULER \\
\enddata
\tablecomments{HJD is the Heliocentric Julian Day --2450000 days.
The goodness of fit of the image, $\chi^2/N_{dof}$, is used to rescale the
formal uncertainties by a factor of $(\chi^2/N_{dof})^{1/2}$.  The Image A1-C
columns give the magnitudes of the quasar images relative to the
comparison stars.}
\label{tab:lceuler}
\end{deluxetable}

\clearpage

\end{document}